\documentclass[
aps,
prx,
amsmath,
amssymb,
aps,
superscriptaddress,
twocolumn,
]{revtex4-2}
\usepackage[T1]{fontenc} 
\usepackage{graphicx}
\usepackage{dcolumn}
\usepackage{bm}
\usepackage{hyperref}
\usepackage{appendix}
\usepackage{xcolor}
\usepackage{braket}

\usepackage[normalem]{ulem} 

\begin{document} 

\title{Protocols for a many-body phase microscope:\\ From coherences and $d$-wave superconductivity to Green's functions}

\author{Christof Weitenberg}
\email {christof.weitenberg@tu-dortmund.de}
\affiliation{Department of Physics, TU Dortmund University, 44227 Dortmund, Germany}

\author{Luca Asteria}
\affiliation{Department of Physics, Graduate School of Science, Kyoto University, Kyoto 606-8502, Japan}

\author{Ola Carlsson}
\affiliation{Department of Physics and Arnold Sommerfeld Center for Theoretical Physics (ASC),
Ludwig-Maximilians-Universität München, 80333 München, Germany}
\affiliation{Munich Center for Quantum Science and Technology (MCQST), 80799 München, Germany}

\author{Annabelle Bohrdt}
\affiliation{Department of Physics and Arnold Sommerfeld Center for Theoretical Physics (ASC),
Ludwig-Maximilians-Universität München, 80333 München, Germany}
\affiliation{Munich Center for Quantum Science and Technology (MCQST), 80799 München, Germany}

\author{Fabian Grusdt}
\affiliation{Department of Physics and Arnold Sommerfeld Center for Theoretical Physics (ASC),
Ludwig-Maximilians-Universität München, 80333 München, Germany}
\affiliation{Munich Center for Quantum Science and Technology (MCQST), 80799 München, Germany}

\begin{abstract}
Quantum gas microscopes probe quantum many-body lattice states via projective measurements in the occupation basis, enabling access to various density and spin correlations. Phase information, however, cannot be directly obtained in these setups. Recent experiments went beyond this by measuring local current operators and local phase fluctuations. Here we propose how Fourier-space manipulation in a matter-wave microscope allows access to various long-range off-diagonal correlators in experimentally realistic settings, realizing a many-body phase microscope. We demonstrate in particular how the fermionic $d$-wave superconducting order parameter in arbitrary Hubbard-type models, the non-equal time Green's function yielding the spectral function, or the hidden order of composite bosons in a fractional Chern insulator can be directly measured. Our results show the great potential of matter-wave microscopy for accessing exotic correlators including phases and coherences and characterizing intriguing quantum many-body states. 
\end{abstract}


\maketitle

\begin{figure}
    \centering
    \includegraphics[width=\linewidth]{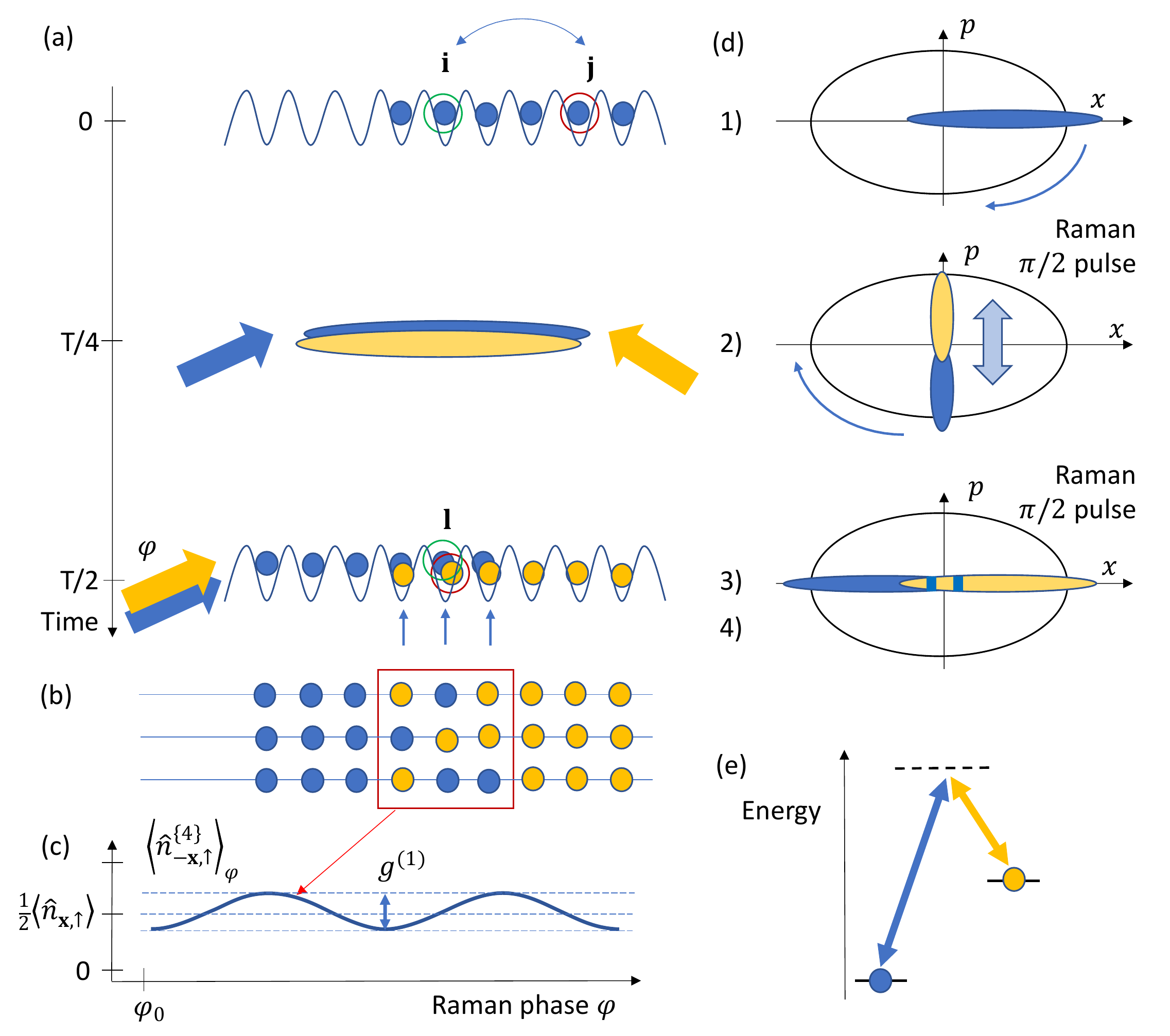}
    \caption{Proposed protocol for measuring equal-time Green's functions and off-diagonal long-range order. (a) Sketch of the general protocol using a matter-wave microscope. Details are described in the main text. 
    1) Atoms are initially in an optical lattice in a strongly-correlated regime, all in spin state $\uparrow$ (blue) ($t=0$). 
    2) After instantaneously switching off the lattice and interactions, a $T/4$ matter-wave pulse transforms the system into Fourier space, where directly afterwards (at $t=T/4$) a $\pi/2$ Raman pulse brings the system into superposition with an auxiliary spin state $\downarrow$ (yellow) that acquires a momentum kick. 
    3) After another $T/4$ step, a second $\pi/2$ Raman pulse in the matter-wave image plane (at $t=T/2$) but without momentum transfer makes lattice sites with a given distance $\mathbf{d}=\mathbf{i}-\mathbf{j}$ interfere (note that the matter-wave protocol inverts the image).
    4) Spin-resolved single-atom imaging. 
    (b) Example results of snapshot measurements.
    (c) If there is coherence between sites $\mathbf{i}$ and $\mathbf{j}$, one expects to see clear fringes in the measured density $\langle \hat{n}_{\mathbf{x},\uparrow}^{\{4\}}  \rangle_\varphi$ of the initial spin state $\uparrow$ as a function of the phase $\varphi$ of the second Raman pulse. Their amplitude corresponds directly to the $g^{(1)}(\mathbf{d})$ correlation function. The phase offset $\varphi_0$ of the fringe (e.g. due to time-reversal symmetry breaking) encodes the complex phase of $g^{(1)}(\mathbf{d})$. (d) Illustration of the protocol in phase space, where $T/4$ pulses swap the roles of $x$ and $p$, indicating the interferometer scheme.  (e) The blue and yellow arrows indicate the utilized Raman transition between the two spin states.}
    \label{fig:1}
\end{figure}

\section{Introduction}
Ultracold atoms in optical lattices constitute a formidable experimental platform for studying quantum many-body systems due to their tunability and versatile detection schemes. In particular, quantum gas microscopes allow projective measurements of the many-body system with single-site resolution and single-atom sensitivity \cite{Bakr2010,Sherson2010,Gross2021} allowing access to density-density correlators \cite{Endres2011} and also multi-point spin correlators \cite{Koepsell2019} or string order correlators \cite{Endres2011,Hilker2017}. Several works proposed protocols for also accessing off-diagonal correlations and coherences \cite{Killi2012,Knap2013,Kessler2014,Kosior2014,Ardila2018,Murthy2019} and recent experiments demonstrated measurements of local current operators \cite{Impertro2024, Impertro2025} and local phases \cite{Bruggenjurgen2024} with single-lattice-site resolution. These measurements ultimately rely on a mapping to density, either mapping local currents to density imbalances using controlled tunneling dynamics in local double wells \cite{Impertro2024} or mapping local phase fluctuations to density fluctuations via a matter-wave analog of phase contrast imaging \cite{Murthy2019,Bruggenjurgen2024}. Related measurements also used spin-resolved dynamics in local double wells to map spin singlet and triplet states to densities~\cite{Greif2013}. Another way of accessing the degree of coherence is Talbot interferometry \cite{Santra2017}, where the contrast of successive Talbot revivals probe the degree of coherence over increasing distances and which was recently also combined with local resolution \cite{Bruggenjurgen2024,Koehn2026,Su2026}.

An intriguing class of ultracold atom experiments is realized by the \textit{matter-wave microscope}~\cite{Asteria2021}. These setups use a sequence of quarter-period evolutions in harmonic traps to map between momentum and position distributions~\cite{Tung2010}, forming time-domain matter-wave analogues to optical lenses. Combining two matter-wave lenses with different trap frequencies yields a magnified matter-wave image of the original density distribution with a magnification given by the ratio of the trap frequencies and magnifications up 93 were realized ~\cite{Asteria2021}. A major advantage of this magnification is that in the subsequent optical imaging effects from diffraction and high densities are suppressed. This concerns both the optical resolution, which then allows to resolve structures below the lattice spacing~\cite{Asteria2021} and the depth of focus, where single-site resolved images of systems that are extended along the perpendicular direction become possible, e.g. arrays of tubes realized in 2D optical lattices~\cite{Zahn2022}. Matter-wave microscopy has also been combined with single-atom imaging, which allows to better resolve strongly-correlated regimes such as the pairing of fermions~\cite{Brandstetter2025}.

A second major advantage of matter-wave microscopes is the access to the matter-wave Fourier space between the two matter-wave lenses, which can be used for manipulation in combination with detection in real space. The new possibilities that arise from this have so far only started to be explored. One possibility is phase imprinting in Fourier space to obtain a matter-wave analog of phase contrast imaging~\cite{Murthy2019}, which was realized by exploiting the effective phase imprinting from matter-wave aberrations~\cite{Bruggenjurgen2024}. Another possibility is mapping of momentum space to an auxiliary dimension to realize a phase-space microscope~\cite{Cooper2026}. Here we focus on the implications of this Fourier space access, and show how it unlocks a powerful new toolbox for the investigation of many-body quantum systems, specifically as it allows measurements of highly non-local off-diagonal correlators. These protocols can be regarded as a many-body version of a phase microscope, because they give access to the relevant correlation functions that characterize strongly-correlated states.

\section{Many-body interferometer} 
\label{secManyBodyInterf}
Here we propose protocols using the matter-wave microscope with Fourier-space manipulation \cite{Murthy2019,Bruggenjurgen2024} for accessing various non-local off-diagonal correlators in quantum gas microscopes. The idea is based on a many-body interferometer scheme as sketched in Fig.~\ref{fig:1}. The protocol starts with a matter-wave lens consisting of a time-evolution in a harmonic trap for a quarter of the trap period $T$, which exchanges position and momentum space, i.e., realizes a Fourier transformation~\cite{Murthy2014,Asteria2021}. The matter-wave lens applies to a single-particle picture and interactions are switched off at the start of the protocol (see Section~\ref{experimental}). Then a 
Raman $\pi/2$ pulse into an auxiliary spin state is applied in this Fourier space and transfers an appropriate momentum $\mathbf{q}$ controlled by the choice of the angle $\theta$ between the two Raman beams. The use of such initially unoccupied auxiliary spin states is an established technique in ultracold atoms~\cite{Stewart2008}. The second matter-wave lens realizes another $T/4$ pulse and converts the momentum transfer into a displacement $\mathbf{d}(\mathbf{q})$ in real space \cite{Asteria2022}. A second Raman pulse without momentum transfer then closes the interferometer sequence and leads to interference between the shifted and unshifted system. The interference fringes in spin-resolved measurements, recorded as a function of the Raman phases~\cite{Cronin2009}, reveal the coherence in the system over the distance given by the chosen shift.

The quantities measured after manipulation by matter-wave lenses and Raman beams can be understood by the corresponding evolution of operators in the Heisenberg picture. We define the creation operator \(\hat{a}^\dagger_{\mathbf{x}, \uparrow /\downarrow}\)  for a fermion or boson at position \(\mathbf{x}\) with spin up/down and the operator \(\hat{a}^\dagger_\mathbf{k, \uparrow /\downarrow}\) for the corresponding momentum 
\begin{equation}
    \hbar\mathbf{k}(\mathbf{x})= m \omega_1 \mathbf{x}
\label{eqkx}    
\end{equation}
determined by the trap frequency $\omega_1$ of the matter-wave lens and the particle mass $m$.
With the initial state polarized to the physical spin \(\uparrow\), a density measurement after closing the interferometer in the fourth step then reads (see Appendix)
\begin{multline}
        \langle\hat{n}_{-\mathbf{x},\uparrow}^{\{4\}}  \rangle = \langle\hat{a}^{\{4\}\dagger }_{-\mathbf{x},\uparrow} \hat{a}_{-\mathbf{x},\uparrow}^{\{4\}}  \rangle\\ =\frac{1}{4} [\langle
        \hat{n}_{\mathbf{x}, \uparrow}\rangle + \langle \hat{n}_{\mathbf{x}+\mathbf{d},\uparrow}\rangle -
        (e^{-i\varphi}\langle \hat{a}_{\mathbf{x} + \mathbf{d}, \uparrow }^\dagger \hat{a}_{\mathbf{x}
        , \uparrow} \rangle + \text{h.c.})],
\end{multline}
where $\hat{a}^{\dagger \{n\}} $ denotes any creation operator after sequence step $n$. Expectation values in the second line are evaluated with respect to the original many-body wavefunction before the sequence. This expression contains the interference term between sites separated by distance $\mathbf{d}$, assumed to be an integer multiple of the lattice constant $a_{\rm lat}$. If the system is additionally invariant under time-reversal and translations, the phase coherence function (or equal-time single-particle Green's function)
\begin{equation}
    g^{(1)}(\mathbf{d}) = \langle \hat{a}_{\mathbf{x} + \mathbf{d}, \uparrow }^\dagger \hat{a}_{\mathbf{x}, \uparrow}\rangle
\end{equation}
is real, and one finds (see Fig.~\ref{fig:1} c)
\begin{align}
    \langle\hat{n}_{-\mathbf{x},\uparrow}^{\{4\}}  \rangle &= \frac{1}{2} \langle
        \hat{n}_{\mathbf{x}\uparrow} \rangle  -
        \frac{1}{2} \cos (\varphi)\langle \hat{a}_{\mathbf{x} + \mathbf{d}, \uparrow }^\dagger \hat{a}_{\mathbf{x}
        , \uparrow}\rangle
        \label{eq:g1meas}
.\end{align}

The phase coherence is given by the fringe contrast, which can be obtained by subtracting two measurements where the Raman phase is set to \(\varphi = 0\) and \(\varphi = \pi\). Evaluating many lattice sites allows extracting average values from a single snapshot. Comparing the mean density in the overlapping region of the system and its shifted copy with the mean density away from the overlap even allows extracting the phase coherence from a single snapshot. 
In the case when \(g^{(1)} \in \mathbb{C}\) is complex, e.g., in the presence of gauge fields, for systems out of equilibrium or with spontaneous symmetry breaking, the fringe will be shifted in \(\varphi\), and mapping out the signal w.r.t.\ the interferometer phase $\varphi$ gives access also to the complex phase of $g^{(1)}$, similar to the phase microscope \cite{Bruggenjurgen2024}.

The idea discussed above is very general and applies to bosons, fermions and spin-mixtures. It can serve as a powerful building block for far more complex measurement sequences. In the following we discuss several concrete protocols, which access particularly relevant correlators for ongoing quantum simulation efforts. We also discuss experimental considerations at the end of the manuscript.

\section{d-wave superconducting order}
\label{secDwaveSC}
The question under which conditions the repulsive Fermi-Hubbard model hosts a $d$-wave superconductor at low temperatures constitutes an urgent quest of quantum-many body physics~\cite{Roth2025}. Recent progress in the preparation of Fermi-Hubbard systems with ultracold atoms, through new cooling techniques~\cite{Xu2025} and the entrance of the pseudogap regime~\cite{Chalopin2024,Kendrick2026}, pushes these systems towards an answer. The $d$-wave order parameter can be accessed through the measurement of pairing correlations, i.e., four-point correlators of the form 
\begin{equation}
C_{\mu,\nu}(\mathbf{i}-\mathbf{j}) = \langle \hat{a}_{\mathbf{i}\uparrow}^\dagger \hat{a}^\dagger_{\mathbf{i}+\mathbf{e}_\mu\downarrow} \hat{a}_{\mathbf{j}\uparrow} \hat{a}_{\mathbf{j}+\mathbf{e}_\nu\downarrow} \rangle
\label{eq:dcorrelator}
,\end{equation}
including displacements along the unit vectors $\mathbf{e}_x$ and $\mathbf{e}_y$ in the $\mu,\nu=x$ and $y$ directions; here $\uparrow, \downarrow$ correspond to two physical spin states. This correlator is the Fourier transform of the momentum-space correlator
\begin{align*}
C_{\mu,\nu}(\mathbf{k}, \mathbf{k}^\prime) &= \langle \hat{a}_{\mathbf{k}\uparrow}^\dagger \hat{a}^\dagger_{-\mathbf{k}\downarrow} \hat{a}^{\vphantom{\dagger}}_{\mathbf{k}^\prime\uparrow} \hat{a}^{\vphantom{\dagger}}_{-\mathbf{k}^\prime \downarrow} \rangle \\
&= \sum_{\mathbf{i}, \mathbf{j}, \mathbf{v}_\mu, \mathbf{v}_\nu}
e^{-i(\mathbf{k}\mathbf{v}_\mu - \mathbf{k}^\prime \mathbf{v}_\nu)}
\langle 
\hat{a}_{\mathbf{i}\uparrow}^\dagger \hat{a}^\dagger_{\mathbf{i}+\mathbf{v}_\mu\downarrow} \hat{a}_{\mathbf{j}\uparrow} \hat{a}_{\mathbf{j}+\mathbf{v}_\nu\downarrow}
\rangle,
\end{align*}
where \(\mathbf{v}_{\mu, \nu}\) are arbitrary lattice vectors. This is the correlator more commonly used in the condensed-matter context, and contains the same information as \eqref{eq:dcorrelator} under the restriction of \(\mathbf{v}_{\mu, \nu}\) to lattice basis vectors.

Next to the preparation of the $d$-wave superconductor, accessing its characteristic long-range off-diagonal pairing correlations $C_{\mu,\nu}(\mathbf{d})$ at large $\mathbf{d}$ is a major challenge and so far only few protocols based on noise correlations have been described \cite{Rey2009,Kitagawa2011}. The problem becomes easier when leveraging the mapping to an attractive Hubbard model~\cite{Ho2009}, where realistic measurement protocols for quantum gas microscopes have been identified \cite{Schlomer2024,Mark2025}. However, the mapping from the repulsive to the attractive Hubbard model fails when the diagonal hopping term $t'$ is included, which has been established to play an important role for stabilizing the $d$-wave phase \cite{Xu2024}. Therefore the call for a feasible measurement protocol of $d$-wave order in the repulsive Fermi Hubbard model is timely.

\begin{figure}
    \centering
    \includegraphics[width=\linewidth]{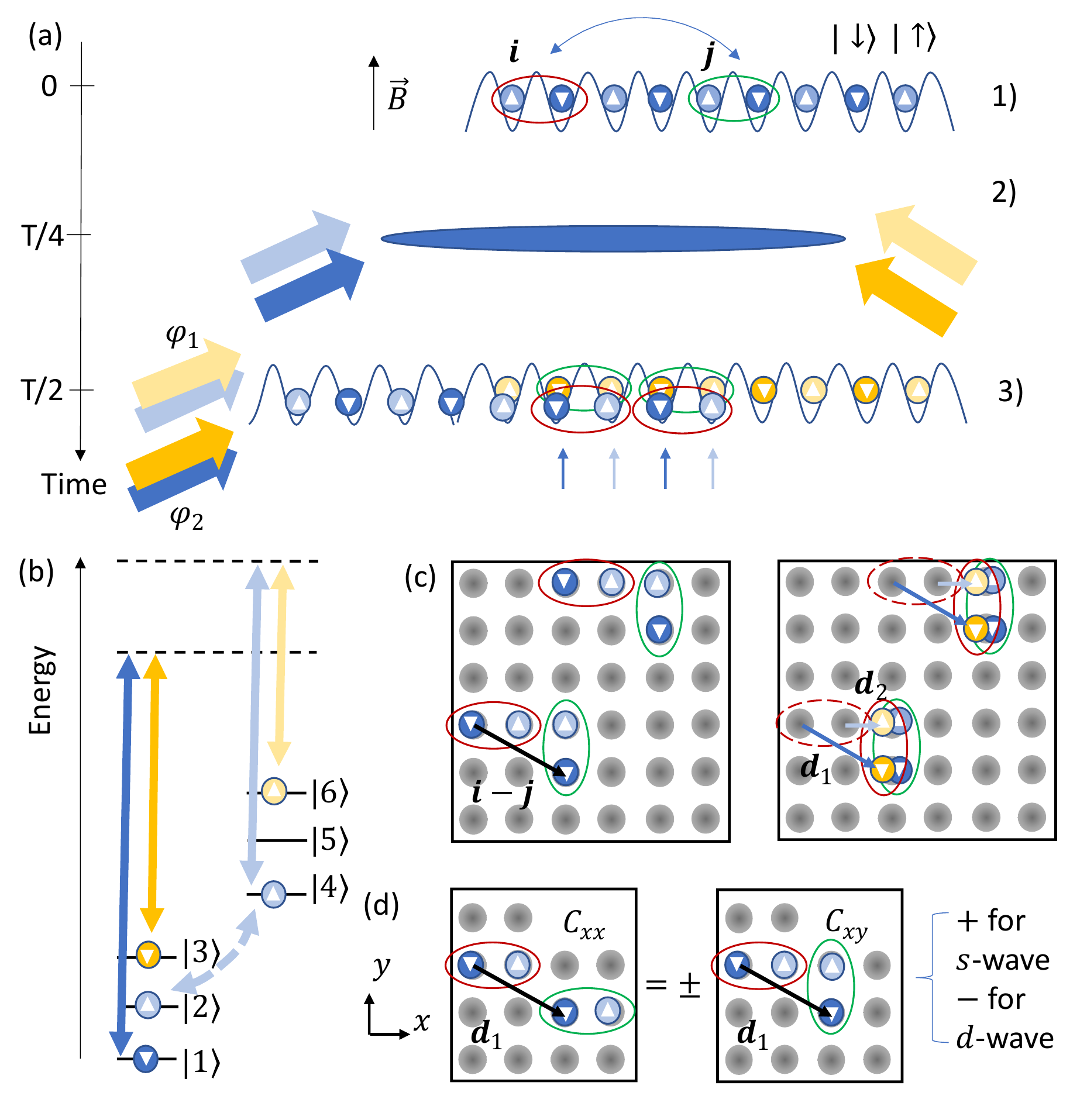}
    \caption{Proposed protocol for measuring the superconducting orders with arbitrary pairing symmetry. (a) Spinful version of the protocol in Fig.~\ref{fig:1} using two physical spins (dark and light blue) and two auxiliary spins (dark and light yellow). Superconducting correlations can be detected from suitable four-point correlators $C_{\mu,\nu}(\mathbf{d})$ (see main text). (b) The four spin states are connected by separate Raman transitions. The spin-up state is transferred from hyperfine state $\ket{2}$ to $\ket{4}$ at the beginning of the protocol (dashed arrow), in order to switch off interactions (see Section~\ref{experimental}). (c) Protocol for distinguishing between different pairing symmetries. The Raman transition for the spin-up state gives a kick that yields a translation by $\mathbf{d}_1=\mathbf{i}-\mathbf{j}$ (dark blue arrow), while the spin-down state is translated by $\mathbf{d}_2=\mathbf{i}-\mathbf{j}+\mathbf{e}_\mu-\mathbf{e}_\nu$ (light blue arrow) -- shown here for the combination $\mu=x$, $\nu=y$. This interferes the wavefunction of a pair of atoms, each displaced along a different direction, with itself at a controlled separation. (d) Illustration of the distinction between $s$-wave and $d$-wave superconducting order.} 
    \label{fig:2}
\end{figure}

Here we show that a spinful version of the general protocol introduced in Sec.~\ref{secManyBodyInterf} allows directly accessing superconducting order in general Hubbard-type models. This is based on the interpretation of $C_{\mu,\nu}(\mathbf{d})$ as a phase coherence function similar to $g^{(1)}(\mathbf{d})$ but formulated for a pair of spins. Thereby the spatial symmetry of the local pair wavefunction is encoded in the dependence on $\mu$ and $\nu$ and allows to further distinguish between different pairing symmetries. E.g., for an $s$-wave ($d$-wave) superconductor $C_{x,y}(\mathbf{d})=+C_{x,x}(\mathbf{d})$ ($C_{x,y}(\mathbf{d})=-C_{x,x}(\mathbf{d})$), i.e. bonds along $\mathbf{e}_x$ and $\mathbf{e}_y$ are in (out-of) phase. Our scheme is summarized in Fig.~\ref{fig:2} and can be applied e.g. to detect $s$-wave ($d$-wave) superconducting order in attractive (repulsive) Hubbard models. 

For the measurement of $C_{\mu,\nu}(\mathbf{d})$, we propose to use two auxiliary spin states complementing the two physical spin states. Performing the same interferometry steps as described in Sec.~\ref{secManyBodyInterf} for the measurement of \(g^{(1)}\), but independently between each physical spin and its corresponding auxiliary spin, see Fig.~\ref{fig:2} a-b, different displacements \(\mathbf{d}_1 = \mathbf{i}-\mathbf{j}
\) and \(\mathbf{d}_2 =\mathbf{i} - \mathbf{j} + \mathbf{e}_\mu - \mathbf{e}_\nu\) can be independently realized for $\uparrow$ and $\downarrow$ respectively by choosing corresponding momentum kicks of the two Raman pulses. Moreover, the phases of the two Raman pulses, serving as control phases in the Ramsey interferometer, can be independently set to \(\varphi_1\), \(\varphi_2\). By choosing other Raman kicks, pairing correlations involving beyond-nearest neighbor pairs $\sim \hat{a}_{\mathbf{i}\uparrow}^\dagger \hat{a}^\dagger_{\mathbf{i}+\mathbf{r}\downarrow}$ with pair sizes $\mathbf{r}$ involving larger integer multiples of $\mathbf{e}_\mu$ would also be accessible. Assuming that the system is invariant under time-reversal, one can show that a combination of four measurements with different Raman phases $\varphi_1=\varphi_2=\varphi$ suffices to extract the pairing correlator from density-density correlations obtained in the last step of the sequence (see Appendix):
\begin{align*} 
        &\langle \hat{n}_{-\mathbf{j}\uparrow}^{\{4\}}  \hat{n}_{-\mathbf{j} - \mathbf{e}_\nu \downarrow}^{\{4\}} \rangle_{ \varphi = 0} + \langle \hat{n}_{-\mathbf{j}\uparrow}^{\{4\}}  \hat{n}_{-\mathbf{j} -\mathbf{e}_\nu \downarrow}^{\{4\}} \rangle_{\varphi = \pi} \\
        &-
        \langle \hat{n}_{-\mathbf{j} \uparrow}^{\{4\}}  \hat{n}_{-\mathbf{j} -\mathbf{e}_\nu \downarrow}^{\{4\}} \rangle_{\varphi = \pi /2} 
        -
\langle \hat{n}_{-\mathbf{j} \uparrow}^{\{4\}}  \hat{n}_{-\mathbf{j} -\mathbf{e}_\nu \downarrow}^{\{4\}} \rangle_{ \varphi = -\pi /2}
        \\&= \frac{1}{2}\langle
        \hat{a}^\dagger_{\mathbf{i} \uparrow}\hat{a}^\dagger_{ \mathbf{i} + \mathbf{e}_\mu \downarrow} \hat{a}^{\vphantom{\dagger}}_{\mathbf{j}
                \uparrow} 
                \hat{a}^{\vphantom{\dagger}}_{\mathbf{j} + \mathbf{e}_\nu
        \downarrow}\rangle = \frac{1}{2} C_{\mu,\nu}(\mathbf{i}-\mathbf{j})
        \label{eq:dmeas}
.\end{align*}
Here we assumed fermionic operators $\hat{a}^{(\dagger)}_{\mathbf{j},\sigma}$, although a similar relation can be obtained for bosons. 

Computing long-range off-diagonal superconducting correlations $C_{\mu,\nu}(\mathbf{d})$ for ground states of 2D Hubbard models is extremely challenging also numerically. Recent works suggest that with diagonal tunneling $t'$ and at zero temperature, these superconducting correlations drop over a few lattice sites to a constant value of $10^{-4}$~\cite{Roth2025}. The algebraic decay expected for finite temperature will require the detection of signals on a similar level. The signal is expected to be quasi isotropic, such that the correlator corresponding to different $\mathbf{i}-\mathbf{j}$ can be averaged. For resolving the predicted small signal of the universal part of the correlator, we expect to require on the order of $10^4$ measurements, but even from the much stronger signal of the non-universal part at shorter distances, one will be able to access the pairing symmetry and to verify the expected $d$-wave nature of the pairing.

\section{Non-equal-time correlations and ARPES spectra} 
\label{secARPES}
The spectral function $A(\mathbf{k},\omega) = - \pi^{-1} {\rm Im} [G(\mathbf{k},\omega)]$ is a central concept for understanding quantum many-body systems and their excitations \cite{Stewart2008}, e.g., for studying mobile dopants in correlated quantum magnets \cite{Brown2020,Bohrdt2020,Morera2024} or polarons in Bose-Einstein condensates \cite{Jorgensen2016,Skou2022}. In solid-state systems, it can be directly accessed via angle-resolved photoemission spectroscopy (ARPES)~\cite{Damascelli2003,Sobota2021}. Likewise, using Raman pulses~\cite{Dao2007,Dao2009} or lattice modulations~\cite{Kollath2006spec,Bohrdt2020} in the linear response regime, $A(\mathbf{k},\omega)$ can be measured in cold atom systems~\cite{Torma2016} and quantum gas microscopes~\cite{Brown2020}, but obtaining the required statistics in the latter represents a major challenge.
In numerical simulations, on the other hand, it is often more convenient to compute the retarded Green's function $G(\mathbf{k}_0,\omega)$ as the Fourier transform of the non-equal-time correlation function $G(\mathbf{k}_0,t)=-i \theta(t) \langle \hat{a}^\dagger_{\mathbf{k}_0}(t)\hat{a}_{\mathbf{k}_0}(0) \rangle$, with the heavyside step function $\theta(t)$ restricting to times $t>0$. This relation of $A$ and $G$ has also been employed in experimental studies of quantum impurities immersed in a bath~\cite{Cetina2016,Ardila2019,Skou2021}, where Ramsey interferometry can be used to directly measure $G(\mathbf{k}_0,t)$~\cite{Knap2013}, without a linear response assumption.
Obtaining non-equal-time correlation functions with full momentum resolution in bulk many-body systems and without (auxiliary) impurities is a major challenge, however. 

\begin{figure}
    \centering
    \includegraphics[width=\linewidth]{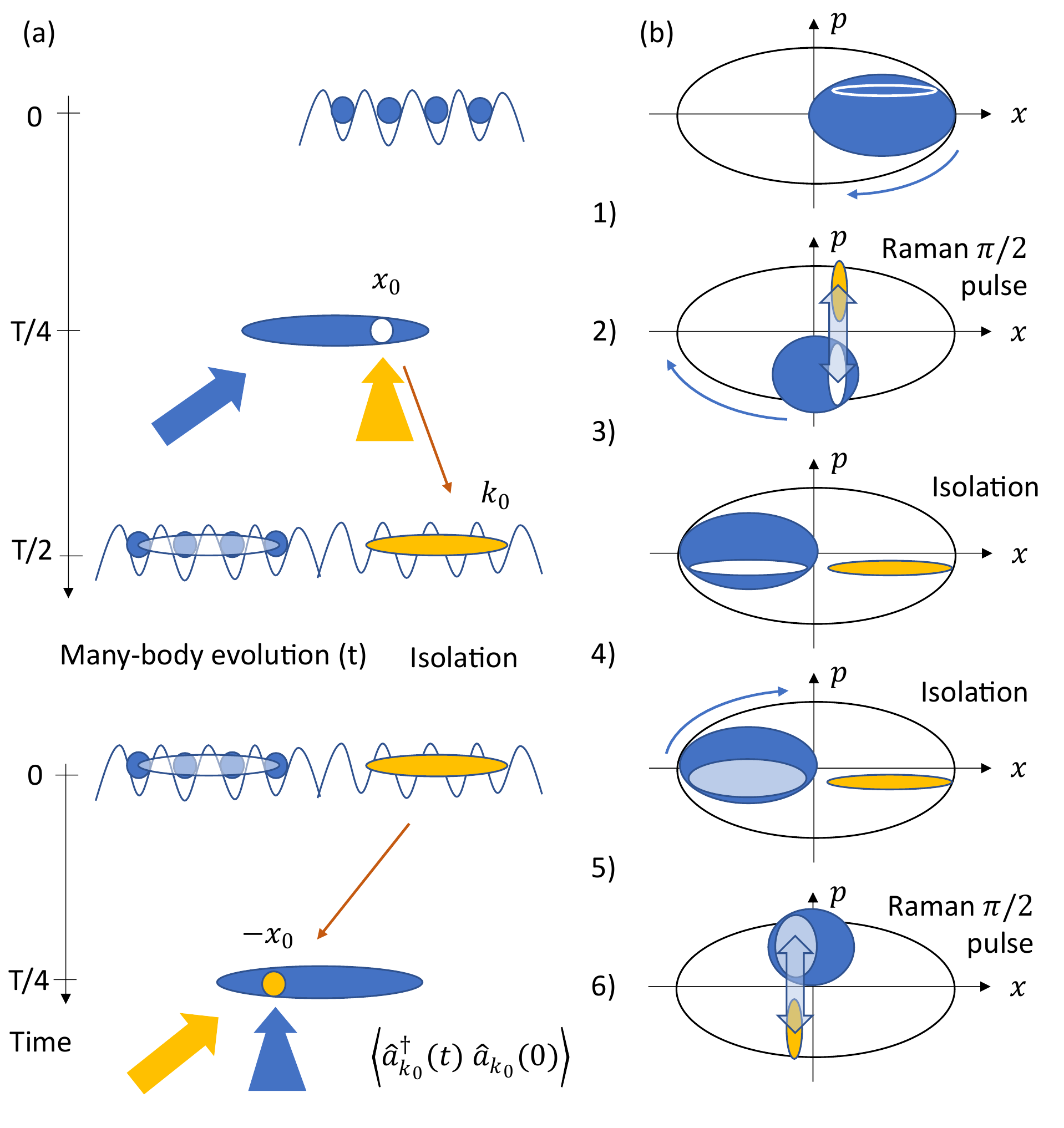}
    \caption{Proposed protocol for measuring non-equal time correlators. (a) 1) A strongly correlated system is initialized, interactions and lattice potentials are switched off, and a $T/4$ pulse takes it into the Fourier plane. 2) A $\pi/2$ Ramsey pulse with one focused Raman beam at $\mathbf{x}_0$ extracts a particle in a selected momentum mode $\mathbf{k}_0(\mathbf{x}_0)$, see Eq.~\eqref{eqkx}. 3) In the image plane (at $T/2$) this mode is shifted and therefore isolated from the remaining many-body system, which 4) undergoes its intrinsic many-body dynamics after reloading into the original lattice and switching on interactions. The extracted momentum mode is indicated by the white hole in the system. 5) In a second Fourier plane (after applying another $T/4$ pulse), the selected momentum mode is brought to interfere with the system via 6) a second $\pi/2$ Raman pulse giving access to the correlator $\langle \hat{a}^\dagger_{\mathbf{k}_0\uparrow}(t)\hat{a}_{\mathbf{k}_0\uparrow}(0) \rangle$ (see main text). (b) Illustration of the protocol in phase space in order to showcase the interferometer sequence.}
    \label{fig:3}
\end{figure}

Here we propose a suitable protocol for directly measuring the retarded Green's function $G(\mathbf{k}_0,t)$ at a pre-defined momentum $\mathbf{k}_0$, based on the matter-wave microscope as summarized in Fig.~\ref{fig:3}. The idea is to extract a particle in momentum mode $\mathbf{k}_0$ from the system, by (partially) transferring it to the auxiliary spin state with a $\pi/2$ Ramsey pulse. This can be achieved using a focused Raman beam in Fourier space, i.e. after a $T/4$ pulse. The extracted particle is then kept in spatial isolation while evolving the rest of the many-body system for a duration $t$. Afterwards, the Ramsey interferometer is closed by returning the extracted particle in a second $\pi/2$ Ramsey pulse. Similar to Fig.~\ref{fig:1}, the Ramsey scheme gives access to an off-diagonal two-point correlator, but now for operators in a single, well-defined momentum mode $\mathbf{k}_0$ and at non-equal times.

In the following we discuss a spin-polarized ($\uparrow$) system coupled to an auxiliary spin state ($\downarrow$). In this case, the $\uparrow$ density measured after completing the final, sixth step of the sequence described above at position $-\mathbf{x}_0$, corresponding to the target momentum $\mathbf{k}_0$ in Fourier space, reads (see Appendix for a detailed derivation):
\begin{multline}
        \langle \hat{n}_{-\mathbf{x}_0 \uparrow}^{\{6\}} \rangle = \frac{1}{2}\langle
        \hat{n}_{\mathbf{k}_0 \uparrow}\rangle \\
        - \frac{1}{4} \left( e^{-i\varphi
        -i\varepsilon_{\mathbf{k}_0}t} \langle \hat{a}^{\dagger}_{ \mathbf{k}_0
\uparrow}(t) \hat{a}_{\mathbf{k}_0 \uparrow}(0) \rangle + \text{h.c.} \right),
     \end{multline}
where the right-hand side of the equation is evaluated with respect to the original many-body wavefunction from the start of the sequence.     
As the non-equal time correlators are generally complex, the fringe contrast must be mapped out over \(\varphi\) to extract the phase of the correlator, \(\varphi_0 = {\rm arg} \langle \hat{a}^{\dagger}_{ \mathbf{k}_0\uparrow}(t) \hat{a}_{\mathbf{k}_0 \uparrow}(0) \rangle\), as sketched in Fig.~\ref{fig:1}(c). 

The selected momentum mode $\mathbf{k}_0$ is spatially isolated from the system during the many-body time evolution, as illustrated in Fig.~\ref{fig:3}. As it typically contains one atom or less, interactions in this mode can be neglected. Further, since the momentum mode is an eigenstate of the lattice, it does not disperse during the many-body time-evolution in the middle of the sequence (see Sec.~\ref{experimental} for further discussion). It only undergoes a dynamical phase evolution $e^{-i \varepsilon_{\mathbf{k}_0}t}$ resulting from the single-particle band structure $\varepsilon_{\mathbf{k}_0}$, which can be calibrated independently, e.g., by a non-interacting reference measurement, and needs to be subtracted from the measured phase of the fringe. In contrast, the many-body system with a particle removed from momentum mode $\mathbf{k}_0$ is no longer an eigenstate of the full many-body Hamiltonian, and its non-trivial time evolution leads to the reduced amplitude in the interference, which produces the desired signal in $G(\mathbf{k}_0,t)$. 

As in the previous protocols (Secs.~\ref{secManyBodyInterf}-\ref{secDwaveSC}) the present scheme involves instantaneously turning off interactions as a first step. However, additional care must be taken with the lattice potential, as the original quasimomentum modes need to eventually be transferred back into the lattice to undergo many-body time-evolution. Instead of instantaneously switching off the lattice, this demands a controlled band mapping (with interactions switched off): By adiabatically lowering the lattice, each quasimomentum state $\boldsymbol{\kappa}$ is transferred into one well-defined momentum state $\mathbf{k}$ of the continuum. Subsequently the desired momentum $\mathbf{k}_0$ can be targeted locally at $\mathbf{x}_0$ in the Fourier plane. Following the interferometry steps in the matter wave microscope, the lattice can be raised again using inverse band mapping, affecting both the many-body system and the outcoupled, i.e. spatially isolated, particle. The latter has a well-defined quasimomentum $\boldsymbol{\kappa}_0$ in the lattice, and the rest of the many-body system transfers to the initial state, albeit with one particle removed in the taget quasimomentum mode $\boldsymbol{\kappa}_0$. Crucially, during band mapping, each single-particle quasimomentum mode $\boldsymbol{\kappa}$ picks up a dynamical phase $\phi(\boldsymbol{\kappa})$, which needs to be compensated in order to coherently restore the system before it can undergo the many-body time evolution. This can be achieved by imprinting a suitable phase in the matter-wave Fourier plane at time $T/4$.

Alternatively, instantaneous ramping of the lattice can still be performed, if some additional precautions are taken. This will identify every occupied quasimomentum state $\boldsymbol{\kappa}$ in the first Brillouin zone (BZ) of the lattice not with a single momentum in the continuum, but a superposition of momenta $\mathbf{k}_{\mathbf{n}} = \boldsymbol{\kappa}+\sum_\mu n_\mu \mathbf{K}_\mu$ (with weights $u_{\mathbf{n}}(\boldsymbol{\kappa})$) interspaced by reciprocal lattice vectors $\mathbf{K}_\mu$, with $\mu=x,y,...$ and $n_\mu \in \mathbb{Z}$. The (continuum) mode $\mathbf{k}_0$ within the BZ, selected in the Fourier plane at $T/4$, therefore only targets extraction of a particle in component $\mathbf{n}=0$ while weight from higher momenta $\mathbf{n}\neq 0$ are not targeted. This corresponds to attempting extraction also of higher bands of the quasimomentum spectrum, but since only the lowest band is assumed occupied in the initial state the signal is simply reduced by a factor $|u_{\mathbf{0}}(\boldsymbol{\kappa})|^2$. The fate of the isolated momentum mode demands some attention however, as loading it instantaneously into the lattice will lead to umklapp processes and non-trivial time evolution. This could be circumvented e.g.\ by loading it into an auxiliary harmonic trap away from the bulk lattice system, where revolutions in a matter wave lens can impart trivial time evolution for this leg of the interferometer.

While spectroscopy relies on the linear-response regime for accessing the spectral function, which requires compromises with the signal strength, this real-time approach avoids this limitation and should allow for measurements with good signal and high spectral resolution. 

\begin{figure}
    \centering
    \includegraphics[width=0.85\linewidth]{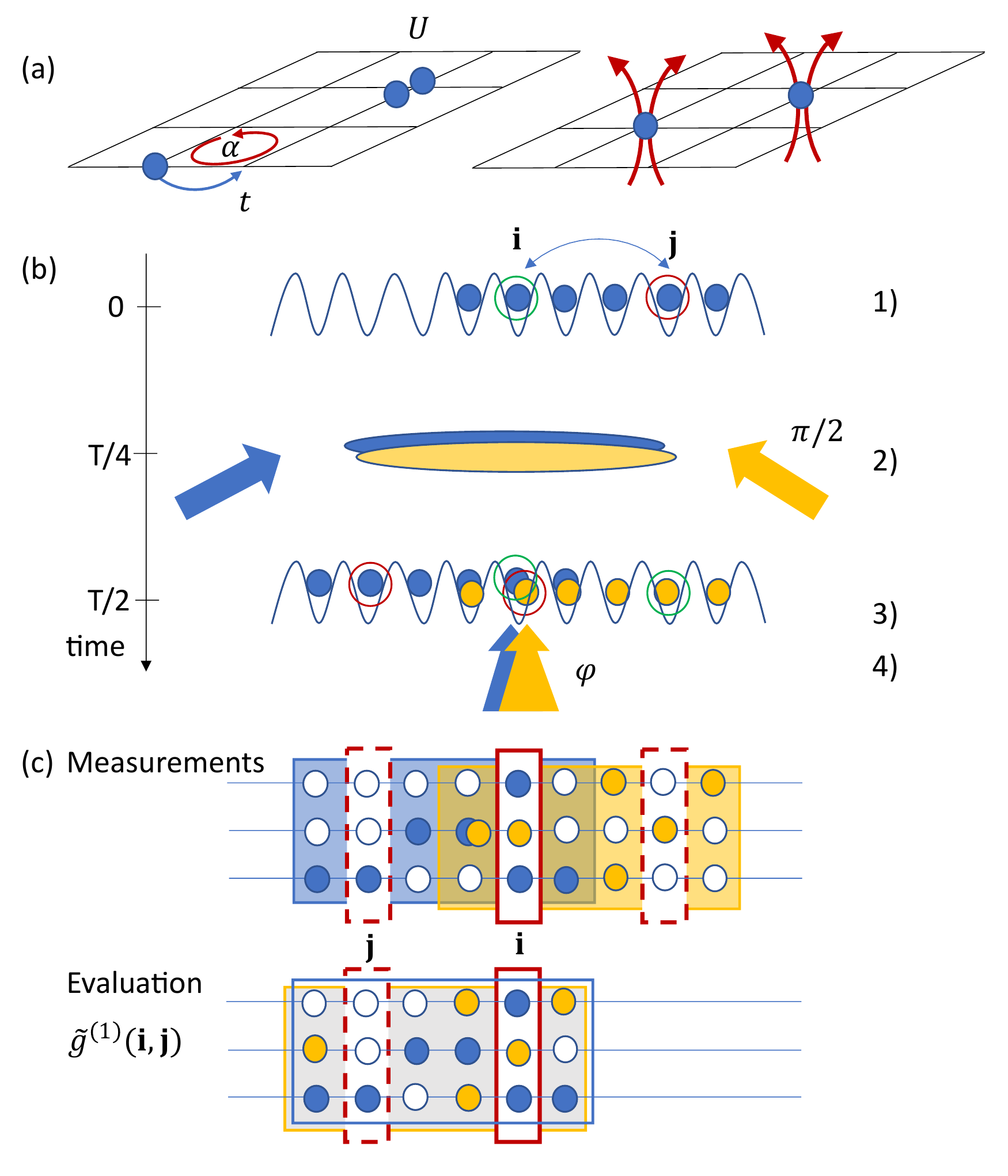}
    \caption{Measuring hidden off-diagonal long-range order in fractional quantum Hall systems. (a) 2D lattice with magnetic flux $\alpha$ and on-site interaction $U$, which gives rise to fractional quantum Hall phases for suitable parameters (left). Such systems can be described by composite bosons by attaching magnetic flux tubes (right). (b) Proposed protocol for measuring off-diagonal long-range correlations of composite bosons. 1)-4) the steps are as in Fig.~\ref{fig:1}, but the second Raman pulse in step 3) is only locally applied via tweezers through a microscope. (c) For the evaluation, the measurements of the auxiliary spin (yellow) are shifted back to obtain the total density (each row is a cut through the 2D system), except for the two sites $\mathbf{i}, \mathbf{j}$ connected by the focused Raman beam (red box), where the coherence is evaluated from the spin contrast. The data in the dashed red boxes does not enter in the final evaluation.}
    \label{fig:4}
\end{figure}

\section{Hidden off-diagonal order}
Another topic at the focus of cold atom quantum simulation efforts are topological phases, which feature intriguing properties such as topologically protected excitations. In recent work with cold atoms in optical potentials, minimal versions of strongly-correlated topological states consisting of up to three interacting bosons~\cite{Leonard2024,Kwan2026} and two interacting fermions~\cite{Lunt2024} were realized, but for larger systems detection of fractional Chern insulators remains an open problem~\cite{Wang2022}. The challenge is rooted in the fact that topological order is by definition non-local and requires the evaluation of observables at extensively many sites simultaneously. While no conventional local order parameter exists in terms of the constituent particles, one can define a conventional order parameter for an effective model of composite particles including a flux attachment~\cite{Read1989}. The fractional quantum Hall phase can be understood as a condensation of such composite bosons, signaled by an off-diagonal long-range order in the $g^{(1)}$ function of the composite particles. 

A recent proposal~\cite{Pauw2024} discussed that this hidden order could be detected with cold atoms by measuring the phase coherence of one particle between two given sites $\boldsymbol{i}, \boldsymbol{j}$ simultaneously with the positions $z_n=x_n+iy_n$ of all other particles, $n=1,...,N-1$. Combining the results in an appropriate correlator~\cite{Read1989,Pauw2024} can then reveal the hidden order underlying the fractional quantum Hall effect. Specifically, the one-particle correlation function for the composite bosons $\tilde{g}^{(1)}(\boldsymbol{i},\boldsymbol{j})$, featuring long-range off-diagonal order in the fractional quantum Hall state, can be expressed in terms of the original annihilation (creation) operators in the form 
\begin{equation} \label{Eq:LatticeHODLRO}
	\tilde{g}^{(1)}(\boldsymbol{i},\boldsymbol{j}) =  \left\langle \prod_{\boldsymbol{l}\neq \boldsymbol{i},\boldsymbol{j}} \left( \frac{z_{\boldsymbol{i}} - z_{\boldsymbol{l}} }{|z_{\boldsymbol{i}} - z_{\boldsymbol{l}}|} \right)^{-\frac{\hat{n}_{\boldsymbol{l}}}{\nu}} \left( \frac{z_{\boldsymbol{j}} - z_{\boldsymbol{l}} }{|z_{\boldsymbol{j}} - z_{\boldsymbol{l}}|} \right)^{\frac{\hat{n}_{\boldsymbol{l}}}{\nu}} \hat{a}^{\dagger}_{\boldsymbol{i}} \hat{a}_{\boldsymbol{j}} \right\rangle. 
\end{equation} 
Here $z_{\boldsymbol{j}} = j_x +\mathrm{i} j_y$ is the position in the complex plane corresponding to site $\mathbf{j}$, and $\nu$ is the filling factor, i.e., the number of particles per flux quantum. This expression shows that the composite boson correlation function picks up an additional non-trivial phase for each occupied lattice site $\boldsymbol{l}\neq \boldsymbol{i},\boldsymbol{j}$ with respect to the bare correlator $g^{(1)}(\boldsymbol{i},\boldsymbol{j})= \langle \hat{a}_{\boldsymbol{i}}^{\dagger}\hat{a}_{\boldsymbol{j}}\rangle$ and this correlation function can be accessed by combining a measurement of the coherence $g^{(1)}(\boldsymbol{i},\boldsymbol{j})$ between sites $\boldsymbol{i}$ and $\boldsymbol{j}$ with a simultaneous measurement of the local occupations $\{\hat{n}_{\boldsymbol{l}}\}_{\boldsymbol{l}\neq \boldsymbol{i},\boldsymbol{j}}$ at all other lattice sites.

The previous work \cite{Pauw2024} suggested a realization of this hybrid measurement using a folded system, which allows connecting sites that are far apart. Here we show that the matter-wave microscope allows accessing this nonlocal order without the need for difficult to realize geometries involving folding the system. The protocol is a variation of the protocol introduced in Fig.~\ref{fig:1}, but replaces the global final Raman pulse by a local pulse, which is focused onto a single lattice site via the microscope objective also used for imaging \cite{Weitenberg2011} (Fig.~\ref{fig:4}). Because the two different spin states at sites $\boldsymbol{l} \neq \boldsymbol{i}, \boldsymbol{j}$ are not coupled at the end of the sequence, they do not interfere and one can reconstruct the original density by summing the signal after moving the shifted component back in space during data evaluation. Adding a local $\pi/2$ Raman pulse at one site $\boldsymbol{j}$ will lead to interference only there, allowing to combine the coherence between two sites with the density at all other sites as required for the correlator of Eq.~(\ref{Eq:LatticeHODLRO}). While measurements according to the protocol in Fig.~\ref{fig:1} would access the phase coherence of the bare particles showing exponential decay, measurement according to Fig.~\ref{fig:4} will access the phase coherence of the flux-attached composite bosons, showing algebraic decay. Based on the numerical study in~\cite{Pauw2024}, we expect that the measurement of the hidden off-diagonal long-range order requires only a few thousand snapshots.

The protocol discussed here mixes measurements of coherences and densities to reveal hidden order in fractional quantum Hall systems. Hidden order has also been revealed in 1D Mott insulators \cite{Endres2011} and in the spin sector for 1D antiferromagnets with hole doping \cite{Hilker2017} and protocols similar to this one might identify so-far unknown hidden order parameters in Hubbard models, which may also include coherences.

\section{Experimental considerations}\label{experimental}
We assume a realization with bosonic $^{87}$Rb atoms for the protocols of Figs.~\ref{fig:1} and \ref{fig:4} and with fermionic $^6$Li atoms for the protocols of Figs.~\ref{fig:2} and \ref{fig:3}. The mass of the respective atom is $m$ and we assume a lattice spacing of $a_{\rm lat}=500$\,nm in both cases. For $^{87}$Rb atoms, the Raman wavelength can be chosen around $\lambda=804~$nm~\cite{Lin2011} and we assume a first matter-wave lens with trap frequency $\omega_1=2\pi \times 50$\,Hz, i.e., a pulse duration of 5~ms. The varying momentum transfer $\boldsymbol{q}$ due to the Raman transition in the Fourier plane will map to a displacement $\boldsymbol{d}=\hbar \boldsymbol{q}/(m \omega_1)$ in the image plane that is an integer multiple of the lattice spacing ($d=n \times a_{\rm lat}$), in order to probe the coherence between lattice sites at varying distance. 

For an angle $\theta$ between the Raman beams, the modulus of the momentum transfer is $ q=(2\pi/\lambda)\sqrt{2(1-\cos(\theta))}$ and a displacement over $n=10$ lattice sites is achieved for an angle of $\theta=16^\circ$ for $^{87}$Rb atoms. The Raman beams can therefore enter through a state-of-the-art high-resolution objective (NA=0.5), which allows an easy change of the angle between the two beams \cite{Ha2015}. For the $^6$Li atoms, we assume a Raman wavelength at $\lambda=671$\,nm  and a trap frequency of $\omega_1=2\pi\times250$\,Hz. In this case, a shift by $n=10$ lattice sites can be realized via an angle between the Raman beams of only $\theta=5^\circ$, which means that an angular precision around $0.05^\circ$ is required for control to a fraction of a lattice site. 
In order to guarantee a stable phase reference, the Raman setup must be interferometrically stable and the beat signal between the two laser beams must be referenced to the start of the Raman pulse. For the second Raman pulse, when no momentum transfer is desired, the two laser beams should be co-propagating, but they can also be provided through the objective. The system remains confined in a 2D plane during the matter wave protocol in order to allow recapture in the lattice.

We assume throughout a matter-wave protocol with unit magnification, which only has the purpose to provide access to Fourier space. The atoms can then be reloaded into the original lattice at the end of the protocol, which can be used to freeze them for single-atom resolved imaging \cite{Bakr2010,Sherson2010,Gross2021}. Spin-resolved single-atom imaging can be realized by splitting the two spin states along the perpendicular direction with suitable lattice manipulations \cite{Koepsell2020}, which will be insensitive to possible in-plane excitations from the recapture in the lattice. Alternatively, very large matter-wave magnifications can be combined with free-space fluorescence imaging \cite{Brandstetter2025}. A larger matter-wave magnification could help with the local Raman addressing in the image plane discussed in Fig.~\ref{fig:4}. 

An important point for realizing the matter-wave microscope is to switch off interactions during the matter-wave protocol, because it relies on a single-particle picture. For the protocols using $^{87}$Rb atoms, the physical spin is $F=1$, $m_F=1$ and the auxiliary spin is $F=2$, $m_F=2$. To keep interaction effects low, we suggest to switch off the vertical confinement to rapidly decrease the density and add a matter-wave condition also in the vertical direction~\cite{Murthy2014}. We note that switching off the lattice already contributes to a substantial reduction of the density and thereby of the interaction effects.

For the $^6$Li atoms, one can use the six hyperfine states of the ground state, which we label $\ket{1}$ to $\ket{6}$ in ascending order of energies. In the Paschen-Back regime, the three lowest hyperfine states $\ket{1}$, $\ket{2}$, and $\ket{3}$ are high-field seeking and the upper three hyperfine states $\ket{4}$, $\ket{5}$ and $\ket{6}$ are low-field seeking. One can choose, e.g., $\ket{1}$ and $\ket{2}$ as the initial physical spin states and a suitable magnetic field to realize a strong interaction via a Feshbach resonance~\cite{Hulet2020}. At the beginning of the protocol, one can then flip one of the spin states to one of the low-field seeking states, e.g. $\ket{2}$ to $\ket{4}$. The transition can be driven very quickly, because it involves flipping the electron spin~\cite{Holten2022}. This constitutes a rapid switch-off of interactions, because there are no Feshbach resonances between the low-field seeking and the high-field seeking states and the scattering length is given by the singlet scattering length $a_s=47 a_0$ \cite{Brandstetter2025}, which is negligible compared to the scattering lengths achieved with the Feshbach resonance. The Raman pulses to the auxiliary spin states can then be chosen within the high-field seeking states and the low-field seeking states, respectively, such that Feshbach resonances are avoided, e.g., by coupling $\ket{1}-\ket{3}$ and $\ket{4}-\ket{6}$ (Fig.~\ref{fig:2}b).

The initial flip of one spin state is applied at the beginning of the protocol together with a switch-off of the optical lattice in order to initiate the protocol. It is not shown in Figs.~\ref{fig:2} and \ref{fig:3} for clarity. These Raman pulses are realized with two co-propagating beams in order to avoid a momentum transfer. 

The duration of all Raman pulses in the protocols need to be short compared to the many-body time scales and compared to the duration of the matter-wave lenses in order to cause negligible displacement. With a trap frequency of $\omega_1=2\pi \times 250$\,Hz, i.e. quarter-period matter-wave lenses of $T=1$\,ms duration, and achievable Raman pulse durations down to 200~ns \cite{Holten2022}, this can be fulfilled. 

In the protocol of Fig.~\ref{fig:3}, we select a momentum mode using a focused Raman beam in the Fourier space of the matter-wave microscope. 
We choose a suitable 1/e$^2$ waist radius of the focused Raman beam $w_0$ and matter-wave trap frequency $\omega_1=2\pi \times250$\,Hz for the mapping from momentum space to real space $x=p/m\omega_1$. In order to select a mode width with typically one atom, we assume a system of 20 sites length and select a mode with 1/20 of the reciprocal lattice vector in momentum space, which corresponds to $w_0=(h/a_{\rm lat}/20)/(m\omega_1)=4 \mu m$ for $^6$Li atoms. This can be tuned either via $w_0$ or $\omega_1$. For this waist size, the spread of wave vectors, which could cause a spread in the Raman momentum transfer, can be neglected.

The loading back into the lattice has to be realized as an inverse band mapping, which maps the isolated momentum mode onto a quasimomentum mode, i.e., an eigenstate of the lattice, which has no dispersion but only a trivial phase evolution of energy \(\varepsilon_{\mathbf{k}_0}\). In (inverse) band mapping, the time scale for ramping the lattice has to be slow compared to the single-particle band gaps, which can be around 100 µs for $^6$Li atoms, which is fast compared to the matter-wave lens duration of $(2\pi/\omega_1)/4=1$\,ms. 

We assume that the original system has a very weak overall confinement or a box potential and that the confinement with trap frequency $\omega_1$ is switched on rapidly for the matter-wave lenses only. It is then not a limitation for the (inverse) band mapping time scale. With an angle of $30^\circ$ between the Raman lasers, the isolated mode is shifted by $n_{\rm iso}=68$ lattice sites. For the many-body evolution, it can then be loaded into an isolated region of the lattice of this size, possibly bounded by an additional box potential. Because the quasi momentum mode is not an exact eigenstate of a finite system, we expect a dephasing with a frequency around $J/n_{\rm iso}$ with the tunneling energy $J$, which sets one (tolerable) limitation on the achievable frequency resolution.
Loading back into the lattice also requires excellent phase stability of the lattice over the duration of the matter-wave protocol and a high-fidelity matter-wave imaging of the original distribution without aberration-induced shifts. Reloading with good matching of the wave function to the lattice was demonstrated after time evolutions on the order of a Talbot time in \cite{Santra2017}. Note that in the case of a lattice with inversion-symmetry breaking such as a superlattice, one should use a $3T/4$ matter wave lens in order to avoid the inverting imaging of the $T/4$ pulse.

We note that the proposed protocols will allow accessing exotic order, for which currently no measurement strategy exists. Therefore, also a measurement with deviations from systematic errors, e.g., introduced by residual interactions during the matter-wave manipulation, would be extremely valuable. A detailed analysis of the systematic errors depends on the initial state under investigation and is beyond the scope of this work. We provide an estimate of errors from interactions during the matter-wave protocols for the worst-case scenario of a fully coherent system in Appendix D and find that errors can be largely suppressed by transversely expanding the system.

\section{Conclusion and outlook}
Our results show the great potential of matter-wave microscopy for accessing exotic correlators including phases and coherences and for characterizing intriguing quantum many-body states. We expect that more protocols for relevant correlators can be found. E.g. extensions of the scheme for non-equal time correlators to negative time evolutions from inverting the sign of every term in the Hamiltonian~\cite{Braun2013} might also give access to out-of-time-order correlators, which provide a proxy for diagnosing chaos in quantum systems such as for distinguishing ergodic from non-ergodic dynamics in the context of many-body localization \cite{Keyserlingk2018,Sierant2025}. Furthermore, phase coherence is an interesting signature for Majorana zero modes at the edges of a 1D topological system, where the two-point correlation function $\langle \hat{a}_1^\dagger \hat{a}_j \rangle$ exponentially decays in the bulk and has a revival at the other end of the chain \cite{Defossez2025}. Similarly, power-law decay of $g^{(1)}$ correlations along edges of fractional quantum Hall states can be probed. It might also be worthwhile to study which further information the correlators discussed here can provide similar to the extraction of transport coefficients from many current measurements as proposed in \cite{Palm2025}. Moreover, the scheme could be extended to continuous (non-lattice) systems. Finally, going beyond the four-point correlators discussed for the superconducting order, the interference of many-body systems might be useful for extracting other relevant quantities such as entanglement entropy as realized in many-body Hong-Ou Mandel starting from two copies of the system \cite{Islam2015}. 

\section{Acknowledgments}
We thank Sandra Brandstetter for useful comments on the manuscript. We acknowledge funding from the Deutsche Forschungsgemeinschaft (DFG, German Research Foundation) via the Research Units FOR 5919 (Project No. 550495627) (A.B. and C.W.) and FOR 5688 (Project No. 521530974) (C.W.) and via Germany's Excellence Strategy -- EXC-2111 -- (Project No. 390814868) (O.C., A.B. and F.G), and from the European Research Council (ERC) under the European Union’s Horizon 2020 research and innovation programme via the ERC Starting Grants SimUcQuam (Grant Agreement No. 948141) (O.C. and F.G.) and QuaQuaMA (Grant Agreement No. 101217531) (A.B.).

\section{Appendix}

\subsection{Heisenberg picture gates}

The precise quantities measured after manipulation by matter-wave lenses and Raman beams can be understood by the corresponding evolution of operators in the Heisenberg picture, where both lenses and pulses can be modeled as single-particle gates. From knowing how arbitrary states transform, we can write down conjugation with the associated time evolution operators \(\mathcal{U}_{\text{MaW}}\) (Matter-wave lenses) and \(\mathcal{U}_{\text{R}}\) (Raman pulses), defined here via their action on the creation operators:
\begin{align*}
        &\mathcal{U}_{\text{MaW}} \hat{a}_{\mathbf{x}\uparrow /\downarrow}^\dagger
        \mathcal{U}_{\text{MaW}}^\dagger = \hat{a}_{-\mathbf{k}\uparrow /\downarrow}^\dagger \\
        &\mathcal{U}_{\text{MaW}} \hat{a}_{\mathbf{k}\uparrow /\downarrow}^\dagger
        \mathcal{U}_{\text{MaW}}^\dagger = \hat{a}_{\mathbf{x}\uparrow /\downarrow}^\dagger \\
        &\mathcal{U}_{\text{R}} 
        \begin{pmatrix}
       \hat{a}_{\mathbf{k}\uparrow}^\dagger
        \\ \hat{a}_{\mathbf{k} + \mathbf{q}\downarrow}^\dagger
        \end{pmatrix}
        \mathcal{U}_{\text{R}}^\dagger = \frac{1}{\sqrt{2} }
        \begin{pmatrix}
            1 & e^{i\varphi}\\
        -e^{-i\varphi} & 1
        \end{pmatrix}
        \begin{pmatrix}
    \hat{a}_{\mathbf{k}\uparrow}^\dagger \\ \hat{a}^{\dagger}_{
        \mathbf{k} + \mathbf{q} \downarrow} 
        \end{pmatrix}\\ 
        &\mathcal{U}_{\text{R,q=0}} 
        \begin{pmatrix}
    \hat{a}_{\mathbf{x}\uparrow}^\dagger
        \\ \hat{a}_{\mathbf{x}\downarrow}^\dagger
        \end{pmatrix}
        \mathcal{U}_{\text{R,q=0}} ^\dagger = \frac{1}{\sqrt{2} }
        \begin{pmatrix}
            1 & e^{i\varphi}\\
        -e^{-i\varphi} & 1
        \end{pmatrix}
        \begin{pmatrix}
        \hat{a}_{\mathbf{x}\uparrow}^\dagger \\ \hat{a}^{\dagger}_{
        \mathbf{x}\downarrow} 
        \end{pmatrix}
.\end{align*}

The time evolutions of operators in the Heisenberg picture are the inverses of this, i.e.\ we make use of
\begin{align*}
       &\mathcal{U}_{\text{MaW}}^\dagger \hat{a}_{\mathbf{k}\uparrow /\downarrow}^\dagger
        \mathcal{U}_{\text{MaW}} = \hat{a}_{-\mathbf{x}\uparrow /\downarrow}^\dagger \\
        &\mathcal{U}_{\text{MaW}}^\dagger \hat{a}_{\mathbf{x}\uparrow /\downarrow}^\dagger
       \mathcal{U}_{\text{MaW}} = \hat{a}_{\mathbf{k}\uparrow /\downarrow}^\dagger \\
        &\mathcal{U}_{\text{R}}^\dagger 
        \begin{pmatrix}
    \hat{a}_{\mathbf{k}\uparrow}^\dagger
        \\ \hat{a}_{\mathbf{k} + \mathbf{q}\downarrow}^\dagger
        \end{pmatrix}
        \mathcal{U}_{\text{R}} = \frac{1}{\sqrt{2} }
        \begin{pmatrix}
            1 & -e^{i\varphi}\\
        e^{-i\varphi} & 1
        \end{pmatrix}
        \begin{pmatrix}
        \hat{a}_{\mathbf{k}\uparrow}^\dagger \\ \hat{a}^{\dagger}_{
        \mathbf{k} + \mathbf{q} \downarrow} 
        \end{pmatrix}\\ 
        &\mathcal{U}_{\text{R,q=0}}^\dagger 
        \begin{pmatrix}       \hat{a}_{\mathbf{x}\uparrow}^\dagger
        \\ \hat{a}_{\mathbf{x}\downarrow}^\dagger
        \end{pmatrix}
        \mathcal{U}_{\text{R,q=0}} = \frac{1}{\sqrt{2} }
        \begin{pmatrix}
            1 & -e^{i\varphi}\\
        e^{-i\varphi} & 1
        \end{pmatrix}
        \begin{pmatrix}
        \hat{a}_{\mathbf{x}\uparrow}^\dagger \\ \hat{a}^{\dagger}_{
        \mathbf{x}\downarrow} 
        \end{pmatrix}
.\end{align*}
These gates can then be applied to relate operators between the interferometer steps:
\begin{align*}
    \hat{a}^{{\{n\}}\dagger}  = \mathcal{U}_n^{\dagger} \hat{a}^{{\{n-1\}}\dagger} \mathcal{U}_n
,\end{align*}
where \(\hat{a}^{\dagger {\{n\}}}\) denotes any creation operator after sequence step \(n\). The initial operators, encoding observables of the initial state, are written without brackets.

For example, following the scheme of Fig.~\ref{fig:1}, applying the initial matter-wave lens maps 
\begin{align*}
    \hat{a}^{\{1\}\dagger}_{\mathbf{k}, \uparrow /\downarrow} &= \hat{a}_{-\mathbf{x}, \uparrow /\downarrow}^\dagger
.\end{align*}
Here steps are enumerated as in Fig.~\ref{fig:1}. 
The Raman pulse of momentum transfer \(\mathbf{q}\) gives the Heisenberg operators after the second step
\begin{align*}
        \begin{pmatrix}
            \hat{a}_{\mathbf{k}, \uparrow}^{\{2\}\dagger} \\
            \hat{a}_{\mathbf{k} + \mathbf{q}, \downarrow}^{\{2\}\dagger}
        \end{pmatrix} &=
        \frac{1}{\sqrt{2} }
        \begin{pmatrix}
            1 & -1 \\ 1 & 1
        \end{pmatrix}
        \begin{pmatrix}
            \hat{a}^\dagger_{-\mathbf{x} \uparrow} \\ 
        \hat{a}^\dagger_{-\mathbf{x} - \mathbf{d} \downarrow}
        \end{pmatrix}
,\end{align*}
where \(\mathbf{d}\) is the real-space kick corresponding to \(\mathbf{q}\), etc.

The composition yields the operators after all interferometer steps in terms of the initial operators:
\begin{align*}
        &\begin{pmatrix}
            \hat{a}^{\{4\}\dagger}_{\mathbf{x}, \uparrow} \\ 
        \hat{a}^{\{4\}\dagger}_{\mathbf{x}, \downarrow}
        \end{pmatrix} =\\
        &\frac{1}{2} \begin{pmatrix}
             \hat{a}^\dagger_{-\mathbf{x}, \uparrow} -
        e^{i\varphi}\hat{a}^\dagger_{-\mathbf{x} +
        \mathbf{d}, \uparrow} - \hat{a}^\dagger_{-\mathbf{x} - \mathbf{d}, \downarrow} -
e^{i\varphi}\hat{a}^\dagger_{-\mathbf{x}, \downarrow} \\
e^{-i\varphi}\hat{a}_{-\mathbf{x}, \uparrow}^\dagger + \hat{a}^\dagger_{-\mathbf{x}+\mathbf{d}, \uparrow} -
e^{-i\varphi} \hat{a}^\dagger_{-\mathbf{x}-\mathbf{d}, \downarrow} + \hat{a}_{-\mathbf{x},
\downarrow}^\dagger
        \end{pmatrix}
.\end{align*}
For a system invariant under time-reversal (complex conjugation), \(g^{(1)} = \langle \hat{a}_{\mathbf{x} + \mathbf{d}, \uparrow }^\dagger \hat{a}_{\mathbf{x}
        , \uparrow}\rangle =  \langle \hat{a}_{\mathbf{x} + \mathbf{d}, \uparrow }^\dagger \hat{a}_{\mathbf{x}
        , \uparrow}\rangle^* = g^{(1)*}\) implies that two-point correlators offdiagonal in space are real. Evaluating the density expectation value for a spin-polarized state then yields Equation~\eqref{eq:g1meas}.

\subsection{$d$-wave superconducting order}

For the interferometer described in Fig.~\ref{fig:2}, each physical spin leg is coupled through the Raman pulses to an auxiliary spin \(\uparrow \sim \Uparrow\), \(\downarrow \sim\Downarrow\). The Raman momentum kicks \(\mathbf{d}_1\), \(\mathbf{d}_2\), and phases \(\varphi_1\), \(\varphi_2\), are also kept separate, such that the result reads
\begin{align*}
        &\begin{pmatrix}
            \hat{a}^{\{4\}\dagger}_{\mathbf{x}_1, \uparrow} \\ 
        \hat{a}^{\{4\}\dagger}_{\mathbf{x}_1, \Uparrow}
        \end{pmatrix} = \\
        &\frac{1}{2} \begin{pmatrix}
             \hat{a}^\dagger_{-\mathbf{x}_1, \uparrow} -
        e^{i\varphi_1}\hat{a}^\dagger_{-\mathbf{x}_1 +
        \mathbf{d}_1, \uparrow} - \hat{a}^\dagger_{-\mathbf{x}_1 - \mathbf{d}_1, \Uparrow} -
e^{i\varphi_1}\hat{a}^\dagger_{-\mathbf{x}_1, \Uparrow} \\
e^{-i\varphi_1}\hat{a}_{-\mathbf{x}_1, \uparrow}^\dagger + \hat{a}^\dagger_{-\mathbf{x}_1+\mathbf{d}_1, \uparrow} -
e^{-i\varphi_1} \hat{a}^\dagger_{-\mathbf{x}_1-\mathbf{d}_1, \Uparrow} + \hat{a}_{-\mathbf{x}_1,
\Uparrow}^\dagger
        \end{pmatrix},\\
        &\begin{pmatrix}
            \hat{a}^{\{4\}\dagger}_{\mathbf{x}_2, \downarrow} \\ 
        \hat{a}^{\{4\}\dagger}_{\mathbf{x}_2, \Downarrow}
        \end{pmatrix} =\\
        &\frac{1}{2} \begin{pmatrix}
             \hat{a}^\dagger_{-\mathbf{x}_2, \downarrow} -
        e^{i\varphi_2}\hat{a}^\dagger_{-\mathbf{x}_2 +
        \mathbf{d}_2, \downarrow} -\hat{a}^\dagger_{-\mathbf{x}_2 - \mathbf{d}_2, \Downarrow} -
 e^{i\varphi_2}\hat{a}^\dagger_{-\mathbf{x}_2, \Downarrow} \\
e^{-i\varphi_2}\hat{a}_{-\mathbf{x}_2, \downarrow}^\dagger + \hat{a}^\dagger_{-\mathbf{x}_2+\mathbf{d}_2, \downarrow} -
e^{-i\varphi_2} \hat{a}^\dagger_{-\mathbf{x}_2-\mathbf{d}_2, \Downarrow} + \hat{a}_{-\mathbf{x}_2,
\Downarrow}^\dagger
        \end{pmatrix}
.\end{align*}
Taking the initial state to be spin polarized then drops all reference to the auxiliary spins. We find the target correlator in Equation~\eqref{eq:dcorrelator} when selecting e.g.\ \(
        \mathbf{x}_1 = - \mathbf{j}\), \(
        \mathbf{d}_1 = \mathbf{i}-\mathbf{j}\), \(
        \mathbf{x}_2 = -\mathbf{j} - \mathbf{e}_\nu\), \(
        \mathbf{d}_2 = \mathbf{i} - \mathbf{j} + \mathbf{e}_\mu - \mathbf{e}_\nu
\), such that density measurements read
\begin{widetext}
\begin{align}
        \langle \hat{n}_{-\mathbf{j} \uparrow}^{\{4\}}\hat{n}_{-\mathbf{j} - \mathbf{e}_y\nu \downarrow}^{\{4\}} \rangle &=
        \frac{1}{2^4}\langle
            (\hat{n}_{\mathbf{j} \uparrow} +
            \hat{n}_{\mathbf{i} \uparrow})
(\hat{n}_{\mathbf{j} + \mathbf{e}_\nu \downarrow} +
            \hat{n}_{\mathbf{i}+\mathbf{e}_\mu \downarrow})
        +  (\hat{n}_{\mathbf{j} \uparrow} +
            \hat{n}_{\mathbf{i}
            \uparrow})(-e^{i\varphi_2}\hat{a}^\dagger_{\mathbf{i} + \mathbf{e}_\mu
            \downarrow} \hat{a}_{\mathbf{j} + \mathbf{e}_\nu \downarrow} + \text{h.c.})
            \nonumber\\ &+
(-e^{i\varphi_1}\hat{a}^\dagger_{\mathbf{i}
            \uparrow} \hat{a}_{\mathbf{j} \uparrow} + \text{h.c.})
(\hat{n}_{\mathbf{j} + \mathbf{e}_\nu \downarrow} +
            \hat{n}_{\mathbf{i}+\mathbf{e}_\mu \downarrow})
    - I\rangle \nonumber\\
         I &= (e^{i\varphi_1}\hat{a}^\dagger_{\mathbf{i}
            \uparrow} \hat{a}^{\vphantom{\dagger}}_{\mathbf{j} \uparrow} + \text{h.c.})(e^{i\varphi_2}\hat{a}^\dagger_{\mathbf{i} + \mathbf{e}_\mu
            \downarrow} \hat{a}^{\vphantom{\dagger}}_{\mathbf{j} + \mathbf{e}_\nu \downarrow} + \text{h.c.})\nonumber\\
          &= e^{i(\varphi_1 + \varphi_2)}\hat{a}^\dagger_{\mathbf{i}
          \uparrow} \hat{a}^{\vphantom{\dagger}}_{\mathbf{j} \uparrow}\hat{a}^\dagger_{\mathbf{i} + \mathbf{e}_\mu
          \downarrow}\hat{a}^{\vphantom{\dagger}}_{\mathbf{j} + \mathbf{e}_\nu \downarrow}  + e^{i(\varphi_1 -
          \varphi_2)}\hat{a}^\dagger_{\mathbf{i} \uparrow} \hat{a}^{\vphantom{\dagger}}_{\mathbf{j}
          \uparrow}\hat{a}^\dagger_{\mathbf{j} + \mathbf{e}_\nu \downarrow}\hat{a}^{\vphantom{\dagger}}_{\mathbf{i} + \mathbf{e}_\mu
          \downarrow} + \text{h.c.}\label{eq:sumofterms}
.\end{align}
\end{widetext}
The evaluation can be greatly simplified by that the desired four-point correlator is expected to be real. Furthermore, the cross-terms of the form $(\hat{n}_{\mathbf{j} \uparrow} + \hat{n}_{\mathbf{i} \uparrow})(-e^{-i\varphi_2}\hat{a}^\dagger_{\mathbf{i} + \mathbf{e}_\mu \downarrow} \hat{a}^{\vphantom{\dagger}}_{\mathbf{j} + \mathbf{e}_\nu \downarrow} + \text{h.c.})$ are real for time reversal symmetric systems, and the desired correlator can be directly obtained from the combination of four measurements with different Raman phases $\varphi_1=\varphi_2=\varphi$. 
\begin{align}
        &\langle \hat{n}_{-\mathbf{j}}^{\{4\}} \hat{n}_{-\mathbf{j} - \mathbf{e}_\nu}^{\{4\}} \rangle_{ \varphi = 0} + \langle \hat{n}_{-\mathbf{j}}^{\{4\}} \hat{n}_{-\mathbf{j} -\mathbf{e}_\nu}^{\{4\}}\rangle_{\varphi = \pi} \nonumber\\
        &-
        \langle \hat{n}_{-\mathbf{j}}^{\{4\}} \hat{n}_{-\mathbf{j} -\mathbf{e}_\nu}^{\{4\}}\rangle_{\varphi = \pi /2} 
        -
\langle \hat{n}_{-\mathbf{j}}^{\{4\}} \hat{n}_{-\mathbf{j} -\mathbf{e}_\nu}^{\{4\}}\rangle_{ \varphi = -\pi /2}
        \nonumber\\&= \frac{1}{2}\langle
        \hat{a}^\dagger_{\mathbf{i} \uparrow} \hat{a}^\dagger_{ \mathbf{i} + \mathbf{e}_\mu \downarrow} \hat{a}^{\vphantom{\dagger}}_{\mathbf{j}
                \uparrow} 
                \hat{a}^{\vphantom{\dagger}}_{\mathbf{j} + \mathbf{e}_\nu
        \downarrow}\rangle = \frac{1}{2} C_{\mu,\nu}(\mathbf{i}-\mathbf{j})\label{eq:dmeas}
.\end{align}
This can be seen as follows. The first term in Eq.~(\ref{eq:sumofterms}) is independent of Raman phases, so the two positive and two negative measurements in Eq.~(\ref{eq:dmeas}) cancel these contributions. The second and third terms contain go as $\cos(\varphi_1)$ and $\cos(\varphi_2)$, so the measurements with Raman phases $\varphi_1=\varphi_2=0$ and $\pi$ cancel these out, while they are zero for the measurements with $\varphi_1=\varphi_2=\pi/2$ and $-\pi/2$. Finally the second term of $I$ cancels out between the measurements with Raman phases $\varphi_1=\varphi_2=\pm\pi/2$ and $\pi$, \(0\), while the first term of \(I\) adds up, between the same pairs of chosen phases, to the desired correlator. As we consider fermions for this protocol, anticommutation of the different spin sectors has here been used in the final step.

\subsection{Non-equal time correlations and ARPES spectra}
For the out-of-time correlation functions we consider the following sequence of gates:
\begin{enumerate}
        \item MaW lens \(
        \begin{pmatrix}
        \hat{a}^{\{1\}\dagger}_{\mathbf{k}\uparrow /\downarrow} \\
                \hat{a}^{\{1\}\dagger}_{\mathbf{x} \uparrow /\downarrow}
                 \end{pmatrix}
                 =
                \begin{pmatrix}
                \hat{a}^\dagger_{-\mathbf{x}
                \uparrow /\downarrow} \\ \hat{a}^\dagger_{\mathbf{k} \uparrow /\downarrow}
               \end{pmatrix}
                \)
        \item Local Raman \begin{align*}
        \begin{pmatrix}
        \hat{a}^{\{2\}\dagger}_{\mathbf{x} \uparrow} \\
                \hat{a}^{\{2\}\dagger}_{\mathbf{x} \downarrow}
                \end{pmatrix}
                =
                &\frac{g(\mathbf{x}-\mathbf{x_0})}{\sqrt{2} }
                \begin{pmatrix}
                1 & -e^{ i \mathbf{q}\mathbf{x}} \\
                e^{- i\mathbf{q}\mathbf{x}} & 1   
                \end{pmatrix}
                \begin{pmatrix}
                \hat{a}^{\{1\}\dagger}_{\mathbf{x}\uparrow} \\
                \hat{a}^{\{1\}\dagger}_{\mathbf{x} \downarrow}
                \end{pmatrix}\\
                &+ \frac{\sqrt{2 - g(\mathbf{x}-\mathbf{x}_0)^2} - g(\mathbf{x}-\mathbf{x}_0)}{\sqrt{2}}\mathbb{I}\begin{pmatrix}
                \hat{a}^{\{1\}\dagger}_{\mathbf{x}\uparrow} \\
                \hat{a}^{\{1\}\dagger}_{\mathbf{x} \downarrow}
                \end{pmatrix}
                \end{align*}
        \item MaW lens \(
        \begin{pmatrix}
        \hat{a}^{\{3\}\dagger}_{\mathbf{k}\uparrow /\downarrow} \\
                \hat{a}^{\{3\}\dagger}_{\mathbf{x} \uparrow /\downarrow}
                \end{pmatrix}
                =
                \begin{pmatrix}              
               \hat{a}^{\{2\}\dagger}_{-\mathbf{x}
                        \uparrow /\downarrow} \\ \hat{a}^{\{2\}\dagger}_{\mathbf{k} \uparrow
                /\downarrow}
                 \end{pmatrix}
                 \)
        \item Time evolution  \begin{align*}\begin{cases}
            \hat{a}^{\{4\}\dagger}_{\mathbf{x}\uparrow /\downarrow} &= \mathfrak{U}(t) \hat{a}^{\{3\}\dagger}_{\mathbf{x}\uparrow /\downarrow}\text{, }\hfill |\mathbf{x}| < |\frac{\mathbf{d}}{2}|\\
            \hat{a}^{\{4\}\dagger}_{\mathbf{x}\uparrow /\downarrow} &= e^{i\hat{H}_0 t} \hat{a}^{\{3\}\dagger}_{\mathbf{x}\uparrow /\downarrow} e^{-i\hat{H}_0 t}\text{, }\hfill |\mathbf{x}| > |\frac{\mathbf{d}}{2}|
        \end{cases}
        \end{align*}
        \item MaW lens \(\begin{pmatrix}
        \hat{a}^{\{5\}\dagger}_{\mathbf{k}\uparrow /\downarrow} \\
                \hat{a}^{\{5\}\dagger}_{\mathbf{x} \uparrow /\downarrow}
                \end{pmatrix}
                =
                \begin{pmatrix}
                    \hat{a}^{\{4\}\dagger}_{-\mathbf{x}
                        \uparrow /\downarrow} \\ \hat{a}^{\{4\}\dagger}_{\mathbf{k} \uparrow
                /\downarrow}
                  \end{pmatrix}\)
        \item Local Raman 
        \begin{align*}
        \begin{pmatrix}
            \hat{a}^{\{6\}\dagger}_{-\mathbf{x} \uparrow} \\
                \hat{a}^{\{6\}\dagger}_{-\mathbf{x} \downarrow} \end{pmatrix} =  \frac{g(\mathbf{x}-\mathbf{x_0})}{\sqrt{2} }
                \begin{pmatrix}1 & -e^{i\varphi + i\mathbf{q}\mathbf{x}} \\
                e^{-i\varphi-i\mathbf{q}\mathbf{x}} & 1   \end{pmatrix}
                \begin{pmatrix}\hat{a}^{\{5\}\dagger}_{-\mathbf{x}\uparrow} \\
                \hat{a}^{\{5\}\dagger}_{-\mathbf{x} \downarrow}
                   \end{pmatrix}&\\
            + \frac{\sqrt{2 - g(\mathbf{x}-\mathbf{x}_0)^2} - g(\mathbf{x}-\mathbf{x}_0)}{\sqrt{2}}\mathbb{I}\begin{pmatrix}
                \hat{a}^{\{5\}\dagger}_{-\mathbf{x}\uparrow} \\
                \hat{a}^{\{5\}\dagger}_{-\mathbf{x} \downarrow}
                \end{pmatrix}&
                \end{align*}
\end{enumerate}
The last Raman pulse could alternatively be applied globally, because the interference is only evaluated locally. 
Here the momentum kicks of the Raman beams have been incorporated through the
phase factors \(e^{i\mathbf{q}\mathbf{x}}\) allowing a complete real-space
representation, and the targeting of a specific mode is carried out by the
envelope \(g(\mathbf{x} - \mathbf{x}_0, w_0/2)\) which falls off at a length scale \(w_0 /2\),
normalized to a value of one at \(\mathbf{x} = \mathbf{x}_0\) (\(w_0\) omitted in many formulae for clarity). The precise shape of this envelope is not crucial to the following argument, but we take it to be of gaussian form \(g(\mathbf{x}, w_0 /2) = e^{-2|\mathbf{x}|^2 / w_0^2}\), such that \(w_0 /2\) is the standard deviation of the envelope. 
Note how the Raman gates are complemented by additional diagonal terms away from \(\mathbf{x} = \mathbf{x}_0\) to ensure unitarity. Contributions due to these drop when measuring at \(-\mathbf{x}_0\).

The time evolution of step (4) follows the fully interacting Hamiltonian \(\hat{H}\), encoded in the superoperator \(\mathfrak{U}(t) \hat{a}^\dagger_k = e^{i\hat{H} t} \hat{a}^\dagger_k e^{-i\hat{H} t}\), for operators with support on the initial extent of the system, here assumed to be \(|\mathbf{x}| < |\frac{\mathbf{d}}{2}|\). Since only a small fraction of the system is removed from this region by the Raman beam (2), the outside of this initial extent \(|\mathbf{x}| > |\frac{\mathbf{d}}{2}|\) is sufficiently dilute for the dynamics to rather be ruled by a non-interacting Hamiltonian \(\hat{H}_0\). This split in the time evolution can be conveniently incorporated by noting that contributions to \(\hat{a}_{\mathbf{x} \uparrow /\downarrow}^{\{4\}} \sim \int d\mathbf{k}^\prime e^{i\mathbf{k}^\prime\mathbf{x}} f_{\mathbf{k}^\prime}\), where
\(f_\mathbf{k} = g(\mathbf{x} - \mathbf{x}_0, w_0 /2)
e^{i\mathbf{q}\mathbf{x}} \hat{a}_{k\uparrow /\downarrow} \), only have support on the
isolated region \(|\mathbf{x}| > |\frac{d}{2}|\) when \(w_0\) is sufficiently broad
with respect to the momentum kick \(\mathbf{q}\).
This follows from
\begin{align*}
&\int d\mathbf{k}^\prime e^{i\mathbf{k}^\prime\mathbf{x}} g(\mathbf{x}^\prime - \mathbf{x}_0, w_0 /2)
e^{i\mathbf{q}\mathbf{x}^\prime} \hat{a}^\dagger_{\mathbf{k}^\prime\uparrow /\downarrow}  
\\
 &=\int d\mathbf{k}^\prime e^{i\mathbf{k}^\prime\mathbf{x}} g((\mathbf{k}^\prime - \mathbf{k}_0), \frac{m\omega_1}{2\hbar}w_0)
e^{i\mathbf{k}^\prime\mathbf{d}} \hat{a}^\dagger_{\mathbf{k}^\prime\uparrow /\downarrow} 
\\
&= (
\tilde{g}(\mathbf{x}, \frac{2\hbar }{m\omega_1} w_0^{-1})e^{i\mathbf{k}_0 \mathbf{x}}) \ast
\hat{a}^\dagger_{\mathbf{x} + \mathbf{d} \uparrow / \downarrow}
.\end{align*}
Here \(\tilde{g}\) denotes the Fourier transform of \(g\), with corresponding inversion of the
characteristic width. E.g. for \(g\) gaussian, this is another gaussian \(\tilde{g}(\mathbf{x}, w_0^{-1} /2) \sim e^{-2|\mathbf{x}|^2/ w_0^{-2}}\). The convolution (\(\ast\)) then ensures that only operators
acting a distance \(\frac{2\hbar }{m\omega_1} w_0^{-1}\) away from \(\mathbf{x} +
\mathbf{d}\) take part in this object. 
With the width \(w_0\) corresponding to the \(e^{-2}\) waist radius \(4\)~\textmu m suggested in Sec.~\ref{experimental}, and with the momentum kick \(\mathbf{q} =\frac{m\omega_1}{\hbar}\mathbf{d} \) on the order of the Raman wavenumber \(2\pi\lambda^{-1}= 2\pi/671 \) nm\(^{-1}\), we find that the relative spread of \(f\) with respect to \(\mathbf{d}\) is a few percent percent, \(\lambda/(\pi w_0)=0.05 \).
The support
of terms arising from \(f_\mathbf{k}\) is then completely delocalized from the initial system location, and for step \((4)\) such terms take on trivial time evolution \(\hat{a}^{\{4\}\dagger}_{\mathbf{k}\uparrow \downarrow} = e^{i\varepsilon_\mathbf{k} t}\hat{a}^{\{3\}\dagger}_{\mathbf{k}\uparrow \downarrow} \),
where \(\varepsilon_\mathbf{k}\) is the dispersion of the non-interacting lattice system.
Remaining terms have support on the initial system location, and follow the interacting time evolution.

With this preliminary in hand, we turn to composing the gates. With \(h(\mathbf{x}-\mathbf{x}_0) = \sqrt{2 - g(\mathbf{x} - \mathbf{x}_0)^2}\) the magnitude of the diagonals of the Raman gates, we find that:
\begin{widetext}
\begin{align*}
        &\begin{pmatrix} 
        \hat{a}^{\{4\}}_{-\mathbf{k} \uparrow} \\
\hat{a}^{\{4\}}_{-\mathbf{k} \downarrow}
\end{pmatrix} =
\frac{1}{\sqrt{2} } \begin{pmatrix} 
        h(\mathbf{x}-\mathbf{x}_0) \mathfrak{U}(t) & - g(\mathbf{x} - \mathbf{x}_0) e^{ i\mathbf{q}\mathbf{x} + i\varepsilon_\mathbf{k}
        t}
        \\
       g(\mathbf{x} - \mathbf{x}_0) e^{ - i\mathbf{q}\mathbf{x} + i \varepsilon_\mathbf{k} t} &
h(\mathbf{x}-\mathbf{x}_0) \mathfrak{U}(t)\end{pmatrix} 
\begin{pmatrix}
\hat{a}_{\mathbf{k} \uparrow} \\
\hat{a}_{\mathbf{k} \downarrow}
\end{pmatrix} 
\\
        &\begin{pmatrix} 
        \hat{a}^{\{6\}}_{-\mathbf{x} \uparrow} \\
\hat{a}^{\{6\}}_{-\mathbf{x} \downarrow}
\end{pmatrix} = \frac{1}{\sqrt{2} }
\begin{pmatrix} h(\mathbf{x}-\mathbf{x}_0) & -g(\mathbf{x} - \mathbf{x}_0)e^{i\varphi +i\mathbf{q}\mathbf{x}} \\
g(\mathbf{x} - \mathbf{x}_0)e^{-i\varphi-i\mathbf{q}\mathbf{x}} & h(\mathbf{x}-\mathbf{x}_0)\end{pmatrix} 
\begin{pmatrix} \hat{a}^{\{4\}}_{-\mathbf{k}\uparrow}  \\
\hat{a}^{\{4\}}_{-\mathbf{k}\downarrow} \end{pmatrix} 
\\
                  &= \frac{1}{2}
                  \begin{pmatrix} 
                          h(\mathbf{x}-\mathbf{x}_0)^2\mathfrak{U}(t) - g(\mathbf{x} - \mathbf{x}_0)^2e^{i\varphi + i \varepsilon_\mathbf{k}t} &
                          -g(\mathbf{x} - \mathbf{x}_0)h(\mathbf{x}-\mathbf{x}_0)(e^{i\mathbf{q}\mathbf{x} +
                          i\varepsilon_\mathbf{k} t} + \mathfrak{U}(t)
                          e^{i\varphi + i\mathbf{q}\mathbf{x}}) \\
                          g(\mathbf{x} - \mathbf{x}_0 h(\mathbf{x}-\mathbf{x}_0)(\mathfrak{U}(t)e^{-i\varphi - i\mathbf{q}\mathbf{x}}+ e^{
                                  i\varepsilon_\mathbf{k}t -
                  i\mathbf{q}\mathbf{x}}) & h(\mathbf{x}-\mathbf{x}_0)^2\mathfrak{U}(t) - g(\mathbf{x} - \mathbf{x}_0)^2e^{-i\varphi +
          i\varepsilon_{\mathbf{k}}t}
                  \end{pmatrix} 
          \begin{pmatrix} \hat{a}_{\mathbf{k}\uparrow} \\
          \hat{a}_{\mathbf{k}\downarrow} \end{pmatrix} 
.\end{align*}
\end{widetext}
The offdiagonal terms drop out as explained above. When measuring at the focus of the Raman beam the Gaussian envelope also no
longer plays any role and we find, with \(\hat{a}(t)\) denoting annihilation operators time-evolved under the interacting Hamiltonian, a fringe proportional to \(G(\mathbf{k}_0, t)=-i\Theta(t)\langle \hat{a}^{\dagger}_{ \mathbf{k}_0 \uparrow}(t) \hat{a}_{\mathbf{k}_0 \uparrow}(0) \rangle\):
\begin{align*}
        \langle \hat{n}_{-\mathbf{x}_0 \uparrow} ^{\{6\}}\rangle &= \frac{1}{4} (\langle
        \hat{n}_{\mathbf{k}_0 \uparrow}\rangle - e^{-i\varphi
        -i\varepsilon_{\mathbf{k}_0}t} \langle \hat{a}^{\dagger}_{ \mathbf{k}_0
\uparrow}(t) \hat{a}_{\mathbf{k}_0 \uparrow} (0) \rangle + \text{h.c.}).
\end{align*}

\subsection{Numerical study of interaction effects}

The protocols proposed give exact results in the absence of interaction effects.
As an estimate of the validity of this approximation, we here numerically simulate the protocol of Fig.~1, i.e.\ the measurement of the off-diagonal correlator \(g^{(1)}(\mathbf{d})\), for a system where interactions are kept on during the matter-wave protocol. As a worst case scenario, we consider a fully coherent bosonic system of \(^{87}\)Rb, where Bragg peaks of the BEC lead to a sharp increase of densities in the Fourier plane, and therefore a sharp increase of interaction effects. It should be stressed that a lower degree of coherence will reduce these densities and therefore the impact of interactions, much like in the fermionic case where densities remain bounded by Pauli exclusion. We restrict ourselves to a one-dimensional lattice for simplicity, an extension to two dimensions should not lead to significant additional aberrations.

We simulate the full time evolution of the condensate wave function \(\Psi_\sigma\) during our applied matter wave lenses and Raman pulses according to the Gross-Pitaevskii equation:
\begin{align*}
    i\hbar \partial_t \Psi_\sigma = -\frac{1}{2m}\nabla^2 \Psi_\sigma + V_{\text{ext}}\Psi_\sigma + \sum_{\sigma^{\prime}} \frac{4\pi\hbar^2a_1}{ml^2} |\Psi_{\sigma^\prime} |^2\Psi_\sigma 
,\end{align*}
where \(\sigma= \uparrow / \downarrow\) indexes the two spin states necessary for the interferometer, and \(m=87\) u is the mass of the bosons. The contact interaction strengths \(\frac{4\pi\hbar^2a_1}{ml^2}\) are parametrized by the s-wave scattering length \(a_1\), and would differ depending on the hyperfine states considered. For \(^{87}\)Rb however we may pick state manifolds where all scattering lengths are of the order \(a_1 \sim 100 a_0\), which we consider for the present numerics. The parameter \(l\) characterizes the transverse length scale of the lattice confinement, matching one-dimensional densities of the condensate wavefunction to three-dimensional densities relevant for scattering. The system is initialized in a lattice ten sites long, with lattice constant \(a_{\text{lat}} =500\) nm, at a depth of \(3E_{\text{rec}}\). The transverse confinement is estimated as the ground state Gaussian width of deep orthogonal lattices at \(\sim 25 E_{\text{rec}}\), that is \(l \sim 100\) nm. The density of particles is fixed by ensuring an average of three particles per lattice site.

The external potential \(V_{\text{ext}}\) is the sum of a harmonic trap of frequency \(\omega_1 = 2\pi \times50\) Hz  constituting the matter-wave lens and the Raman pulses which we model by a potential of the form
\begin{align*}
    V_\text{Raman}(\mathbf{q}) = \Omega\begin{pmatrix}
        0 & e^{-i\mathbf{qx}} \\ e^{i\mathbf{qx}} & 0
    \end{pmatrix}
,\end{align*}
at a strength corresponding to the pulse area \(\Omega \tau = \pi/2\). This acts on the spinor of the condensate \(\Psi = \begin{pmatrix}
    \Psi_\uparrow \\ \Psi_\downarrow
\end{pmatrix}\).

The interferometry steps are integrated using the split-operator method, on a grid of \(N = 10^5\) spatial points spanning \(200\) \textmu m,
to avoid any boundary effects during the matter wave lens operation. We here adopt integer arguments to index the wavefunction after each interferometer step. In particular, the system is initialized into \(\Psi^{\{0\}}\). The first matter-wave lens brings the system into the Fourier plane, state \(\Psi^{\{1\}}\), before the first Raman pulse separates the interferometer legs in \(\Psi^{\{2\}}\). 
Another matter-wave lens takes the system to the state \(\Psi^{\{3\}}\), before the interferometer is closed by the second Raman pulse into state \(\Psi^{\{4\}}\). The measured densities are subtracted between the choices of Raman phase \(\varphi=\pi\), \(\varphi=0\), which constitutes the measured signal. We note that the interactions do not lead to a shift of the effective quarter-period time for the matter-wave imaging condition as observed for large BECs~\cite{Asteria2021}.

\begin{figure}
    \centering
    \includegraphics[width=1\linewidth]{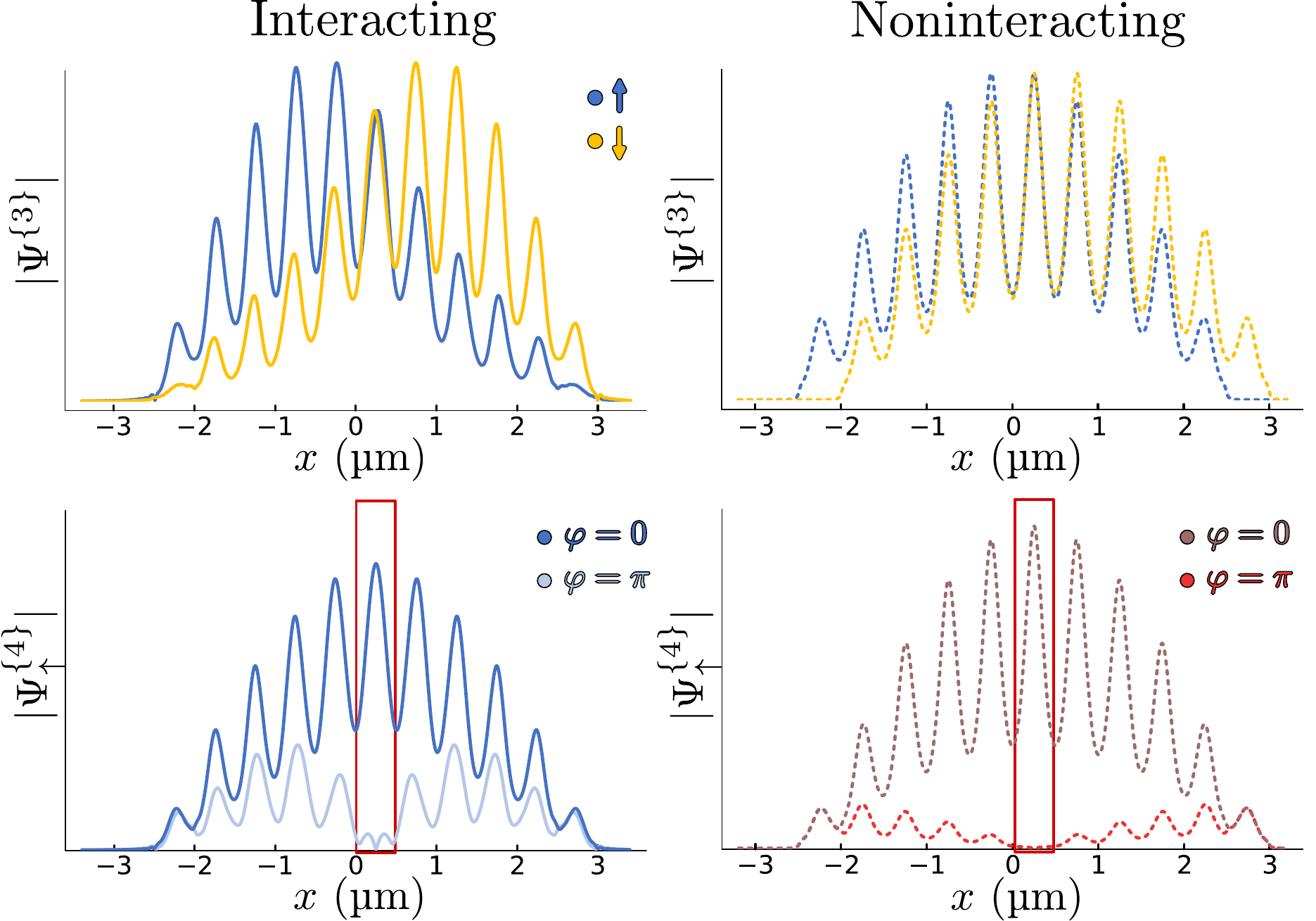}
    \caption{Visual comparison of wavefunctions after various interferometer steps, indexed by integer arguments, between interacting and non-interacting systems. The signal is evaluated as the difference of squared amplitudes between Raman phases \(\varphi=\pi\) and \(\varphi =0\), and the error is estimated within the central Wigner-Seitz cell indicated by the red box.}
    \label{fig:error_comp}
\end{figure}

The errors accumulated due to interactions are quantified by comparing the signal at a single site close to the center of the system with the signal expected from the initial wavefunction. Both of these quantities are averaged over the chosen Wigner-Seitz cell, marked in dashed lines in Figure~\ref{fig:error_comp}, such that the error reads:
\begin{align*}
E = \frac{|\int_{WS} d\mathbf{x} \langle a^\dagger_\mathbf{x} a_{\mathbf{x} + \mathbf{a}}\rangle_\text{m} - 
\int_{WS} d\mathbf{x}
\langle a^\dagger_\mathbf{x} a_{\mathbf{x} + \mathbf{a}}\rangle_\text{e}|}{| \int_{WS} d\mathbf{x} \langle a^\dagger_\mathbf{x} a_{\mathbf{x} + \mathbf{a}}\rangle_\text{e}|}
,\end{align*}
where subscripts m and e are the measured and expected correlation functions, respectively, and \(WS\) denotes the chosen Wigner-Seitz cell. For this error estimate we consider measuring the correlation between neighbouring lattice sites, \(\mathbf{d} = a_{\text{lat}}\).

To reduce interaction effects we let the system evolve in a harmonic trap also in one transverse direction, as described in Section~(VI), at the same frequency as the matter-wave lens trap. This leads to an expansion in the Fourier plane, reducing interaction effects as the Bragg peaks appear, but still yielding a satisfied imaging condition at the end of the protocol, to ensure staying within the depth of focus of a subsequent optical imaging. We model this expansion as the known closed solution for a gaussian wavefunction, the ground state of the transverse confinement before the quench, in an harmonic trap. Such a state remains a gaussian centered in the middle of the trap, but with a standard deviation oscillating with twice the trap frequency \(\omega_1\). Identifying the standard deviation of the gaussian with the transverse length scale \(l\), we find the time-dependence as:
\begin{align*}
    l(t) &= l(0) \sqrt{\cos^2 (\omega_1 t) + l_t^4/l(0)^4 \times \sin^2(\omega_1 t)}
,\end{align*}
where \(l_t = \sqrt{\hbar/m\omega_1}\) is the length scale of the harmonic trap. The transverse length in one direction is then scaled by this factor throughout the protocol, with a maximum magnification of \(l_t^2/l(0)^2 \approx 225\) for a trap frequency of \(50 \) Hz. Note that a similar expansion could also be carried out for a two-dimensional system, as only a single transverse direction is expanded.

The estimated error due to finite interactions is low, \(E \approx 0.9 \%\). The accumulation of errors during the protocol can be further visualized by comparison of the wave function magnitudes during the time evolution, and the measured signal, between the interacting and non-interacting cases, see Figure~\ref{fig:error_comp}. As visible when comparing the left hand side panels to the right hand side, interactions noticeably change the amplitude of each spin species in the central region, leading to a reduced contrast in the signal. The accumulated error however remains small, as the peak density expectation values do not significantly influence the expected spin densities averaged over the Wigner-Seitz cell. We conclude that the transverse expansion during the matter-wave protocol can reduce the density enough to suppress interaction effects in small systems. The fully coherent system considered here provides an upper bound on the expected effect for strongly-correlated initial states with typically short-range coherence.


%

\end{document}